\documentclass[12pt,titlepage]{article}
\usepackage{graphicx}
\def\bbuildrel#1_#2^#3{\mathrel{\mathop{\kern 0pt#1}\limits_{#2}^{#3}}}
\def\slash#1{\setbox0=\hbox{$#1$}#1\hskip-\wd0\dimen0=5pt\advance
       \dimen0 by-\ht0\advance\dimen0 by\dp0\lower0.5\dimen0\hbox
         to\wd0{\hss\sl/\/\hss}}

\newcommand{\m}{\hspace{-1mm}-}
\newcommand{\scb}{\scriptstyle}
\newcommand{\scs}{\scriptscriptstyle}
\newcommand{\be}{\begin{equation}}
\newcommand{\ee}{\end{equation}}
\newcommand{\bea}{\begin{eqnarray}}
\newcommand{\eea}{\end{eqnarray}}
\newcommand{\f}{\frac}
\newcommand{\s}{\hat{s}}

\newcommand{\al}{\alpha_s}
\newcommand{\e}{\epsilon}
\newcommand{\newton}[2]{ \left( \begin{array}{c} 
   {\scb #1} \\ {\scb #2} \end{array} \right) }
\begin{document}
\begin{titlepage}

 \begin{flushright}
  {\bf TUM-HEP-329/98\\       
       IFT-16/98\\
       CERN-TH/99-287\\
      hep-ph/9910220\\
}
 \end{flushright}

 \begin{center}
  \vspace{0.6in}

\setlength {\baselineskip}{0.3in}
{\bf \Large Photonic penguins at two loops and 
$m_t$-dependence of $BR[ B \to X_s l^+ l^-]$}
\vspace{2cm} \\
\setlength {\baselineskip}{0.2in}

{\large  Christoph Bobeth$^{^{1}}$, 
         Miko{\l}aj Misiak$^{^{2,3,4}}$
         and J{\"o}rg Urban$^{^{1,2}}$}\\

\vspace{0.2in}
$^{^{1}}${\it Institut f{\"u}r Theoretische Physik, Technische Universit{\"a}t Dresden, \\
                        Mommsenstr. 13, D-01062 Dresden, Germany}

\vspace{0.2in}
$^{^{2}}${\it Physik Department, Technische Universit{\"a}t M{\"u}nchen,\\
                         D-85748 Garching, Germany}

\vspace{0.2in}
$^{^{3}}${\it Institute of Theoretical Physics, Warsaw University,\\
                 Ho\.za 69, PL-00-681 Warsaw, Poland}

\vspace{0.2in}
$^{^{4}}${\it Theory Division, CERN, CH-1211 Geneva 23, Switzerland}

\vspace{3cm} 
{\bf Abstract \\} 
\end{center} 
\setlength{\baselineskip}{0.3in} 

We calculate two-loop matching conditions for all the operators that
are relevant to \linebreak $B \to X_s l^+ l^-$ decay in the Standard
Model. In effect, we are able to remove the $\pm 16\%$ uncertainty in
the decay spectrum, which was mainly due to the renormalization-scale
dependence of the top-quark mass.  We find $1.46\times10^{-6}$ for the
branching ratio integrated in the domain $m_{l^+l^-}^2/m_b^2 \in
[0.05,0.25]$, for $l= e$~or~$\mu$. There remains around 13\%
perturbative uncertainty in this quantity, while the non-perturbative
effects are expected to be smaller.

\end{titlepage} 

\setlength{\baselineskip}{0.3in}

\section{Introduction}
\label{intro}

The forthcoming measurement of the inclusive decay mode $B \to X_s l^+
l^-$ is expected to provide an important test of possible new physics
effects at the electroweak scale. However, the existing theoretical
predictions for the branching ratio in the Standard Model (SM) still
suffer from many uncertainties, some of which are larger than the
expected experimental errors.

The most important theoretical uncertainties are due to intermediate
$c\bar{c}$ states. Because of the non-perturbative nature of these
states, the differential decay spectrum can be only roughly estimated
when the invariant mass of the lepton pair $m_{l^+l^-}^2$ is not
significantly below $m_{J/\psi}$. It remains questionable whether
integrating the decay rate over this domain can reduce the theoretical
uncertainty below $\pm 20\%$ \cite{LW96}.

On the contrary, for low $\s = m_{l^+l^-}^2/m_{b,pole}^2$ (accessible
to $l = e$~or~$\mu$), a relatively precise determination of the decay
spectrum is possible using perturbative methods only, up to calculable
HQET corrections. The dominant HQET corrections were evaluated in
refs.~\cite{FLS94}--\cite{BI98} and found to be small (smaller than
6\% for $0.05 < \s < 0.25$). Effects of similar size are found in this
region when purely perturbative expressions for $c\bar{c}$
contributions are compared with the ones obtained via dispersion
relations in the factorization approximation (see fig.~\ref{hzs} in
section~\ref{phenom}). Thus, the $B \to X_s l^+ l^-$ decay rate
integrated over this region of $\s$ should be perturbatively
predictable as precisely as the $B \to X_s \gamma$ decay rate, i.e. up
to about 10\% uncertainty.

Unfortunately, the presently available perturbative calculations
\cite{M93,BM95} have not yet reached this precision, even though they
are performed at the next-to-leading (NLO) order in QCD. The formally
leading-order term is (quite accidentally) suppressed, which makes it
as small as some of the NLO contributions. Consequently, some of the
formally next-to-next-to-leading (NNLO) terms can have an effect
larger than 10\% on the differential decay rate. It can be easily
verified by varying the renormalization scale at which the top quark
mass is renormalized in the formulae of refs.~\cite{M93,BM95}.

The formalism of effective theories, which is conventionally used in
the analyses of weak $B$ decays, allows the identification of three
types of NNLO contributions to $B \to X_s l^+ l^-$. The first type
originates from two-loop matching between the Standard Model and the
effective theory amplitudes, i.e.  to two-loop contributions to the
Wilson coefficients in the effective theory at the scale $\mu_0 \sim
M_W$.  The second type is due to the three-loop renormalization group
evolution of the Wilson coefficients down to the scale $\mu_b \sim
m_b$.  The third type originates from two-loop matrix elements of the
effective theory operators between the physical states of interest.
One should include one-loop Bremsstrahlung corrections as well.
Performing a complete NNLO calculation is thus a very involved task.

In the present paper, we shall calculate only the first type of
corrections, i.e. those originating from the two-loop matching
conditions. Our results will allow us to remove the significant
uncertainty of the former NLO prediction stemming from the dependence
on the scale $\mu_0$. The remaining uncalculated NNLO effects will be
estimated in section~\ref{phenom}.

Our paper is organized as follows. In section~\ref{matching}, we
introduce the effective theory and present a complete set of the
matching conditions up to two loops. The resulting formulae for the
so-called effective coefficients are given in
section~\ref{coefficients}. Section~\ref{phenom} is devoted to
discussing phenomenological implications of our results for $B \to X_s
l^+ l^-$.  Technical details of the matching computation are relegated
to section~\ref{details}. There, one can find an extensive description
of the two-loop matching procedure for the photonic penguin diagrams,
which has been the most involved original part of our calculation.
Section~\ref{details} can serve as a practical guide for performing
any two-loop matching computation, not necessarily in the domain of
flavour physics.

\section{Summary of the two-loop matching conditions}
\label{matching}

The effective theory lagrangian relevant to $B \to X_s l^+ l^-$ decay
has the following form
\bea 
{\cal L}_{eff} &=& {\cal L}_{\scs QCD \times QED}(u,d,s,c,b,e,\mu,\tau) 
\nonumber\\
&+& \f{4 G_F}{\sqrt{2}} 
 [  V^*_{us} V_{ub} (C^c_1 P^u_1 + C^c_2 P^u_2) 
  + V^*_{cs} V_{cb} (C^c_1 P^c_1 + C^c_2 P^c_2)]
\nonumber\\
&+& \f{4 G_F}{\sqrt{2}} \sum_{i=3}^{10}  
[(V^*_{us} V_{ub} + V^*_{cs} V_{cb}) C^c_i \; + \; V^*_{ts} V_{tb} C^t_i] P_i. 
\label{Leff}
\eea
For further convenience, we refrain from using unitarity of the CKM
matrix $\hat{V}$ in all the analytical formulae here. The first term
in eq.~(\ref{Leff}) consists of kinetic terms of the light SM
particles as well as their QCD and QED interactions. The remaining two
terms consist of $\Delta B = - \Delta S= 1$ local operators of
dimension $\leq 6$, built out of those light fields:\footnote{
  The $s$-quark mass is neglected here, i.e. it is assumed
  to be negligibly small when compared to $m_b$. Of course, no such
  assumption is made concerning $m_c$ or $m_{\tau}$.}
\be \label{physical}
\begin{array}{rl}
P^u_1 = & (\bar{s}_L \gamma_{\mu} T^a u_L) (\bar{u}_L \gamma^{\mu} T^a b_L), 
\vspace{0.2cm} \\
P^u_2 = & (\bar{s}_L \gamma_{\mu}     u_L) (\bar{u}_L \gamma^{\mu}     b_L),
\vspace{0.2cm} \\
P^c_1 = & (\bar{s}_L \gamma_{\mu} T^a c_L) (\bar{c}_L \gamma^{\mu} T^a b_L),
\vspace{0.2cm} \\
P^c_2 = & (\bar{s}_L \gamma_{\mu}     c_L) (\bar{c}_L \gamma^{\mu}     b_L),
\vspace{0.2cm} \\
P_3 = & (\bar{s}_L \gamma_{\mu}     b_L) \sum_q (\bar{q}\gamma^{\mu}     q),     
\vspace{0.2cm} \\
P_4 = & (\bar{s}_L \gamma_{\mu} T^a b_L) \sum_q (\bar{q}\gamma^{\mu} T^a q),    
\vspace{0.2cm} \\
P_5 = & (\bar{s}_L \gamma_{\mu_1}
                   \gamma_{\mu_2}
                   \gamma_{\mu_3}    b_L)\sum_q (\bar{q} \gamma^{\mu_1} 
                                                         \gamma^{\mu_2}
                                                         \gamma^{\mu_3}     q),     
\vspace{0.2cm} \\
P_6 = & (\bar{s}_L \gamma_{\mu_1}
                   \gamma_{\mu_2}
                   \gamma_{\mu_3} T^a b_L)\sum_q (\bar{q} \gamma^{\mu_1} 
                                                          \gamma^{\mu_2}
                                                          \gamma^{\mu_3} T^a q),
\vspace{0.2cm} \\
P_7  = &  \f{e}{g^2} m_b (\bar{s}_L \sigma^{\mu \nu}     b_R) F_{\mu \nu},
\vspace{0.2cm} \\
P_8  = &  \f{1}{g} m_b (\bar{s}_L \sigma^{\mu \nu} T^a b_R) G_{\mu \nu}^a,
\vspace{0.2cm} \\
P_9  = &  \f{e^2}{g^2} (\bar{s}_L \gamma_{\mu} b_L) \sum_l 
                                      (\bar{l}\gamma^{\mu} l),
\vspace{0.2cm} \\
P_{10} = & \f{e^2}{g^2} (\bar{s}_L \gamma_{\mu} b_L) \sum_l 
                             (\bar{l} \gamma^{\mu} \gamma_5 l),  
\end{array}
\ee
where sums over $q$ and $l$ denote sums over all the light quarks and
all the leptons, respectively. 

        The Wilson coefficients can be perturbatively expanded as
follows
\be \label{expanded.coeffs}
C^Q_i = C^{Q(0)}_i 
+ \f{g^2}{(4 \pi)^2} C^{Q(1)}_i  
+ \f{g^4}{(4 \pi)^4} C^{Q(2)}_i  
+ {\cal O}(g^6), \hspace{1cm} Q = c {\rm ~~or~~} t.
\ee
Their values are found in the matching procedure, which amounts to
requiring equality of $b \to s+$(light particles) Green functions
calculated in the effective theory and in the full Standard Model, up
to ${\cal O}[($external momenta and light masses$)^2/M_W^2]$.
Contributions of order $g^{2n}$ to each Wilson coefficient originate
from $n$-loop SM diagrams, which follows from the particular
convention for powers of gauge couplings in the normalization of our
operators.

Dimensional regularization with fully anticommuting $\gamma_5$ has
been used in our matching computation. Using this simple scheme could
not cause any difficulties, because the choice of the four-quark
operator basis in eq.~(\ref{physical}) allowed us to avoid the
appearance of Dirac traces containing $\gamma_5$ in the effective
theory diagrams \cite{CMM98.df}.  No such traces were present in the
SM diagrams, either.

The $\overline{MS}$ scheme with scale $\mu_0 \sim M_W$ was used
for all the QCD counterterms, both in 
\linebreak \newpage \noindent
the SM and in the effective theory.\footnote{
  The only exceptions were the top-quark-loop contributions to the
  renormalization of the light-quark and gluon wave functions on the
  SM side. The corresponding terms in the propagators were subtracted
  in the MOM scheme at $q^2 = 0$. In consequence, no top-quark loop
  contribution remained in the (W-boson)--(light quark) effective
  vertex after renormalization.}
In addition, several non-physical operators had to be included on the
effective theory side, because the calculation was performed off-shell
(see section~\ref{details} and the appendix for details).

The 't~Hooft--Feynman version of the background field gauge was used
for all the gauge bosons. It allowed us to perform the matching
without making use of the CKM-matrix unitarity.

The only relevant off-shell electroweak counterterm (on the SM side)
proportional to $\bar{s} \slash D b$ was taken in the MOM scheme, at
$q^2 = 0$ for the $\bar{s} \slash \partial b$ term, and at vanishing
external momenta for the terms containing gauge bosons. 

The obtained matching conditions are the following. At the tree level, 
all the $C^{Q(0)}_i$ vanish, except for  
\be \label{c02}
C^{c(0)}_2 = -1.
\ee
The one- and two-loop matching conditions are summarized below:
\bea 
\begin{array}{lclclcl}
C^{c(1)}_1 &=& -15 - 6 L,
& \hspace{3mm} & &&  \\[2mm] 
C^{c(1)}_2 &=& 0, &&&&  \\[2mm] 
C^{c(1)}_3 &=& 0, && 
C^{t(1)}_3 &=& 0,  \\[2mm] 
C^{c(1)}_4 &=& \f{7}{9} - \f{2}{3} L, &&
C^{t(1)}_4 &=& E^t_0(x),  \\[2mm] 
C^{c(1)}_5 &=& 0, && 
C^{t(1)}_5 &=& 0,  \\[2mm] 
C^{c(1)}_6 &=& 0, && 
C^{t(1)}_6 &=& 0,  \\[2mm] 
C^{c(1)}_7 &=& \f{23}{36}, &&
C^{t(1)}_7 &=& -\f{1}{2} A^t_0(x),  \\[2mm] 
C^{c(1)}_8 &=& \f{1}{3}, &&   
C^{t(1)}_8 &=& -\f{1}{2} F^t_0(x),  \\[2mm] 
C^{c(1)}_9 &=& -\f{1}{4 s_w^2} -\f{38}{27} +\f{4}{9} L, &&
C^{t(1)}_9 &=& \f{1-4s_w^2}{s_w^2} C^t_0(x)- \f{1}{s_w^2} B^t_0(x)-D^t_0(x),  \\[2mm]
C^{c(1)}_{10} &=& \f{1}{4 s_w^2}, && 
C^{t(1)}_{10} &=& \f{1}{s_w^2} \left[ B^t_0(x)-C^t_0(x) \right],   \\[2mm]
&&&&&& \\[2mm]
C^{c(2)}_1 &=& T(x)-\f{7987}{72} -\f{17}{3} \pi^2 -\f{475}{6} L - 17 L^2, 
&&&& \\[2mm]
C^{c(2)}_2 &=& -\f{127}{18} -\f{4}{3} \pi^2 -\f{46}{3} L - 4 L^2, &&&& \\[2mm]
C^{c(2)}_3 &=& \f{680}{243} + \f{20}{81} \pi^2 +\f{68}{81} L  + \f{20}{27} L^2, &&
C^{t(2)}_3 &=& G^t_1(x),  \\[2mm]
C^{c(2)}_4 &=& -\f{950}{243} -\f{10}{81} \pi^2 -\f{124}{27} L -\f{10}{27} L^2, && 
C^{t(2)}_4 &=& E^t_1(x),  \\
&&&&&& \\[-2mm]
\end{array} \nonumber
\eea
\bea 
\begin{array}{lclclcl}
C^{c(2)}_5 &=& -\f{68}{243} - \f{2}{81} \pi^2 -\f{14}{81} L - \f{2}{27} L^2, &&
C^{t(2)}_5 &=& -\f{1}{10} G^t_1(x) + \f{2}{15} E^t_0(x),  \\[2mm]
C^{c(2)}_6 &=& -\f{85}{162} -\f{5}{108} \pi^2 -\f{35}{108} L - \f{5}{36} L^2, &&
C^{t(2)}_6 &=& -\f{3}{16} G^t_1(x) + \f{1}{4} E^t_0(x),  \\[2mm]
C^{c(2)}_7 &=& -\f{713}{243} - \f{4}{81} L, &&
C^{t(2)}_7 &=& -\f{1}{2} A^t_1(x),  \\[2mm] 
C^{c(2)}_8 &=& -\f{91}{324}  + \f{4}{27} L, &&
C^{t(2)}_8 &=& -\f{1}{2} F^t_1(x),  \\[2mm] 
C^{c(2)}_9 &=& -\f{1}{s_w^2} -\f{524}{729} +\f{128}{243}\pi^2 +\f{16}{3} L +\f{128}{81} L^2, &&
C^{t(2)}_9 &=& \f{1-4s_w^2}{s_w^2} C^t_1(x)-\f{1}{s_w^2} B^t_1(x,-\f{1}{2}) -D^t_1(x), 
 \\[2mm]
C^{c(2)}_{10} &=& \f{1}{s_w^2}, &&
C^{t(2)}_{10} &=& \f{1}{s_w^2} \left[ B^t_1(x,-\f{1}{2})-C^t_1(x) \right],
\end{array} \nonumber
\eea
where 
\bea
\hspace{2cm}
x = \left( \f{m_t^{\overline{MS}}(\mu_0)}{M_W} \right)^2,
\hspace{2cm}
L = \ln \f{\mu_0^2}{M_W^2},
\hspace{2cm}
s_w = \sin \theta_w
\hspace{23mm}
\eea
and
\bea
\begin{array}{lclr}
A^t_0(x) &=& \f{-3x^3+2x^2}{2(1-x)^4} \ln x + \f{22x^3-153x^2+159x-46}{36(1-x)^3},
&(\arabic{equation}) \addtocounter{equation}{1}\\[4mm]
B^t_0(x) &=& \f{x}{4(1-x)^2} \ln x +\f{1}{4(1-x)},
&(\arabic{equation}) \addtocounter{equation}{1}\\[4mm]
C^t_0(x) &=& \f{3x^2+2x}{8(1-x)^2} \ln x + \f{-x^2+6x}{8(1-x)},
&(\arabic{equation}) \addtocounter{equation}{1}\\[4mm]
D^t_0(x) &=& \f{-3x^4+30x^3-54x^2+32x-8}{18(1-x)^4} \ln x + \f{-47x^3+237x^2-312x+104}{108(1-x)^3},
&(\arabic{equation}) \addtocounter{equation}{1}\\[4mm]
E^t_0(x) &=& \f{-9x^2+16x-4}{6(1-x)^4} \ln x + \f{-7x^3-21x^2+42x+4}{36(1-x)^3},
&(\arabic{equation}) \addtocounter{equation}{1}\\[4mm]
F^t_0(x) &=& \f{3x^2}{2(1-x)^4} \ln x + \f{5x^3-9x^2+30x-8}{12(1-x)^3},
&(\arabic{equation}) \addtocounter{equation}{1}\\[4mm]
A^t_1(x) &=& \f{32x^4+244x^3-160x^2+16x}{9(1-x)^4} Li_2\left(1-\f{1}{x}\right) 
+ \f{-774x^4-2826x^3+1994x^2-130x+8}{81(1-x)^5} \ln x 
&\\[4mm] &&
+ \f{-94x^4-18665x^3+20682x^2-9113x+2006}{243(1-x)^4}
&\\[4mm] &&
+ \left[   \f{-12x^4-92x^3+56x^2}{3(1-x)^5} \ln x
         + \f{-68x^4-202x^3-804x^2+794x-152}{27(1-x)^4} \right] \ln \f{\mu_0^2}{m_t^2},
&(\arabic{equation}) \addtocounter{equation}{1}\\[4mm]
B^t_1(x,-\f{1}{2}) &=& \f{-2x}{(1-x)^2} Li_2\left(1-\f{1}{x}\right) 
+\f{-x^2+17x}{3(1-x)^3} \ln x 
+\f{13x+3}{3(1-x)^2} 
+\left[ \f{2x^2+2x}{(1-x)^3} \ln x + \f{4x}{(1-x)^2} \right] \ln \f{\mu_0^2}{m_t^2}, 
&(\arabic{equation}) \addtocounter{equation}{1}\\[4mm]
C^t_1(x) &=& \f{-x^3-4x}{(1-x)^2} Li_2\left(1-\f{1}{x}\right) 
+ \f{3x^3+14x^2+23x}{3(1-x)^3} \ln x 
+ \f{4x^3+7x^2+29x}{3(1-x)^2} 
&\\[4mm] &&
+ \left[ \f{8x^2+2x}{(1-x)^3} \ln x + \f{x^3+x^2+8x}{(1-x)^2} \right] \ln \f{\mu_0^2}{m_t^2}, 
&(\arabic{equation}) \addtocounter{equation}{1}\\[4mm]
D^t_1(x) &=& \f{380x^4-1352x^3+1656x^2-784x+256}{81(1-x)^4} Li_2\left(1-\f{1}{x}\right) 
+ \f{304x^4+1716x^3-4644x^2+2768x-720}{81(1-x)^5} \ln x 
&\\[4mm] &&
+\f{-6175x^4+41608x^3-66723x^2+33106x-7000}{729(1-x)^4} 
&\\[4mm] &&
+ \left[ \f{648x^4-720x^3-232x^2-160x+32}{81(1-x)^5} \ln x
        +\f{-352x^4+4912x^3-8280x^2+3304x-880}{243(1-x)^4} \right] \ln \f{\mu_0^2}{m_t^2}, 
&(\arabic{equation}) \addtocounter{equation}{1}\\[4mm]
E^t_1(x) &=& \f{515x^4-614x^3-81x^2-190x+40}{54(1-x)^4} Li_2\left(1-\f{1}{x}\right) 
+ \f{-1030x^4+435x^3+1373x^2+1950x-424}{108(1-x)^5} \ln x 
&\\[4mm] &&
+ \f{-29467x^4+45604x^3-30237x^2+66532x-10960}{1944(1-x)^4} 
\end{array}
\nonumber 
\eea
\bea
\begin{array}{lclr}
&&
+ \left[ \f{-1125x^3+1685x^2+380x-76}{54(1-x)^5} \ln x
+ \f{133x^4-2758x^3-2061x^2+11522x-1652}{324(1-x)^4} \right] \ln \f{\mu_0^2}{m_t^2}, 
&(\arabic{equation}) \addtocounter{equation}{1}\\[4mm]
F^t_1(x) &=& \f{4x^4-40x^3-41x^2-x}{3(1-x)^4} Li_2\left(1-\f{1}{x}\right) 
+\f{-144x^4+3177x^3+3661x^2+250x-32}{108(1-x)^5} \ln x
&\\[4mm] &&
+\f{-247x^4+11890x^3+31779x^2-2966x+1016}{648(1-x)^4} 
&\\[4mm] &&
+ \left[ \f{17x^3+31x^2}{(1-x)^5} \ln x
       + \f{-35x^4+170x^3+447x^2+338x-56}{18(1-x)^4} \right] \ln \f{\mu_0^2}{m_t^2}, 
&(\arabic{equation}) \addtocounter{equation}{1}\\[4mm]
G^t_1(x) &=& \f{10x^4-100x^3+30x^2+160x-40}{27(1-x)^4} Li_2\left(1-\f{1}{x}\right) 
+ \f{30x^3-42x^2-332x+68}{81(1-x)^4} \ln x 
&\\[4mm] &&
+\f{-6x^3-293x^2+161x+42}{81(1-x)^3} 
+ \left[ \f{90x^2-160x+40}{27(1-x)^4} \ln x 
       + \f{35x^3+105x^2-210x-20}{81(1-x)^3} \right] \ln \f{\mu_0^2}{m_t^2},
&(\arabic{equation}) \addtocounter{equation}{1}\\[4mm]
T(x) &=& -(16x+8)\sqrt{4x-1} \; Cl_2\left(2 \arcsin \f{1}{2\sqrt{x}}\right) 
+\left(16x+\f{20}{3}\right) \ln x + 32x + \f{112}{9}.
& \hspace{9mm} (\arabic{equation}) \addtocounter{equation}{1}\\
&&&\\[-2mm]
\end{array}
\nonumber 
\eea
The integral representations for the functions $Li_2$ and $Cl_2$ are as follows:
\bea
Li_2(z) &=& -\int_0^z dt \f{\ln (1-t)}{t},\\
Cl_2(x) &=& {\rm Im}\left[ Li_2(e^{ix}) \right] = -\int_0^x d\theta \ln |2 \sin(\theta/2)|.
\eea

Our matching results for all the $C_k^{Q(2)}$ are new, except for
$k=7,8$~and~$10$.  In the cases $k=7$ and $k=8$, we agree with the
previously published results \cite{2mtch}. The $k=10$ case has already
been discussed by us in ref.~\cite{MU99}, and the original calculation
\cite{BB93} has been corrected in ref.~\cite{BB99}.

\section{The effective coefficients}
\label{coefficients}

Once the matching conditions are found, the Wilson coefficients should
be evolved from $\mu_0 \sim M_W$~ to ~$\mu_b \sim m_b$, according to
the Renormalization Group Equation (RGE)
\be \label{RGE}
\mu \f{d}{d\mu} \vec{C}^Q = \left(\hat{\gamma}^Q\right)^T \vec{C}^Q,
\ee
which has the following general solution
\be 
\vec{C}^Q(\mu_b) = \hat{U}^Q(\mu_b,\mu_0) \vec{C}^Q(\mu_0), 
\ee
where \vspace{-1mm}
\bea
\hat{U}^Q(\mu_b,\mu_0) &=& T_g \exp \int_{g(\mu_0)}^{g(\mu_b)} dg' \f{(\hat{\gamma}^Q(g'))^T}{\beta(g')}
\nonumber\\[2mm] && \hspace{-1cm} 
=\hat{U}^{Q(0)}(\mu_b,\mu_0) ~+~ \f{\al(\mu_0)  }{ 4\pi   } \hat{U}^{Q(1)}(\mu_b,\mu_0) 
                                ~+~ \f{\al(\mu_0)^2}{(4\pi)^2} \hat{U}^{Q(2)}(\mu_b,\mu_0) + ...\;.
\eea
In the intermediate step of the above equation, $T_g$ denotes ordering
of the coupling constants such that they increase from right to left.

The anomalous dimension matrices $\hat{\gamma}^Q$ have the following
perturbative expansion
\be \label{gammaexp}
\hat{\gamma}^Q
= \f{\al}{4\pi} \hat{\gamma}^{Q(0)}
+ \f{\al^2}{(4\pi)^2} \hat{\gamma}^{Q(1)}
+ \f{\al^3}{(4\pi)^3} \hat{\gamma}^{Q(2)} + ...\;.
\ee
The one- and two-loop anomalous dimension matrices have already been 
evaluated in refs.~\cite{M93,BM95}. However, transforming them to the
``new'' operator basis (\ref{physical}) is quite non-trivial (see
ref.~\cite{CMM98.df} for the $6 \times 6$ submatrix).  In the ``new''
basis (and in the $\overline{MS}$ scheme with the evanescent operators
specified in the appendix), the matrices $\hat{\gamma}^{c(0)}$ and
$\hat{\gamma}^{c(1)}$ read\footnote
{ Note that the matrices given here correspond to the normalization of
  operators $P_7$, ..., $P_{10}$ as in eq.~(\ref{physical}) and to their
  ordinary Wilson coefficients, not to the so-called ``effective''
  ones that will be introduced below.}
\bea
\hat{\gamma}^{c(0)} &=&
\left(
\begin{array}{cccccccccc}
-4& \f{8}{3}&           0&   -\f{2}{9}&         0&         0&                  0&                  0& -\f{32}{27}&         0\\[2mm] 
12&        0&           0&    \f{4}{3}&         0&         0&                  0&                  0&   -\f{8}{9}&         0\\[2mm] 
 0&        0&           0&  -\f{52}{3}&         0&         2&                  0&                  0&  -\f{16}{9}&         0\\[2mm] 
 0&        0&  -\f{40}{9}& -\f{100}{9}&  \f{4}{9}&  \f{5}{6}&                  0&                  0&  \f{32}{27}&         0\\[2mm] 
 0&        0&           0& -\f{256}{3}&         0&        20&                  0&                  0& -\f{112}{9}&         0\\[2mm] 
 0&        0& -\f{256}{9}&   \f{56}{9}& \f{40}{9}& -\f{2}{3}&                  0&                  0& \f{512}{27}&         0\\[2mm] 
 0&        0&           0&           0&         0&         0& \f{32}{3}-2\beta_0&                  0&           0&         0\\[2mm]
 0&        0&           0&           0&         0&         0&         -\f{32}{9}& \f{28}{3}-2\beta_0&           0&         0\\[2mm]
 0&        0&           0&           0&         0&         0&                  0&                  0&   -2\beta_0&         0\\[2mm]
 0&        0&           0&           0&         0&         0&                  0&                  0&           0& -2\beta_0\\[2mm]
\end{array}
\right),\\[4mm]
\hat{\gamma}^{c(1)} &=&
\nonumber\\[2mm] && \hspace{-2cm}
\left(
\begin{array}{cccccccccc}
-\f{355}{9}& -\f{502}{27}&  -\f{1412}{243}&  -\f{1369}{243}&    \f{134}{243}&   -\f{35}{162}&         -\f{232}{243}&          \f{167}{162}&  -\f{2272}{729}&  0\\[2mm] 
 -\f{35}{3}&   -\f{28}{3}&    -\f{416}{81}&    \f{1280}{81}&      \f{56}{81}&     \f{35}{27}&           \f{464}{81}&            \f{76}{27}&   \f{1952}{243}&  0\\[2mm] 
          0&            0&   -\f{4468}{81}&  -\f{31469}{81}&     \f{400}{81}&  \f{3373}{108}&            \f{64}{81}&           \f{368}{27}&  -\f{6752}{243}&  0\\[2mm] 
          0&            0&  -\f{8158}{243}& -\f{59399}{243}&    \f{269}{486}& \f{12899}{648}&         -\f{200}{243}&        -\f{1409}{162}&  -\f{2192}{729}&  0\\[2mm] 
          0&            0& -\f{251680}{81}& -\f{128648}{81}&   \f{23836}{81}&   \f{6106}{27}&         -\f{6464}{81}&         \f{13052}{27}& -\f{84032}{243}&  0\\[2mm] 
          0&            0&  \f{58640}{243}& -\f{26348}{243}& -\f{14324}{243}& -\f{2551}{162}&       -\f{11408}{243}&         -\f{2740}{81}& -\f{37856}{729}&  0\\[2mm]
          0&            0&               0&               0&               0&              0& \f{4688}{27}-2\beta_1&                     0&               0&  0\\[2mm]
          0&            0&               0&               0&               0&              0&         -\f{2192}{81}& \f{4063}{27}-2\beta_1&               0&  0\\[2mm]
          0&            0&               0&               0&               0&              0&                     0&                     0&       -2\beta_1&  0\\[2mm]
          0&            0&               0&               0&               0&              0&                     0&                     0&               0& -2\beta_1
\end{array}
\right), \hspace{1cm}\\ \nonumber
\eea
where $\beta_0 = \f{23}{3}$ and $\beta_1 = \f{116}{3}$. The analogous
matrices $\hat{\gamma}^{t(0)}$ and $\hat{\gamma}^{t(1)}$ can be
obtained from the ones above by removing the first two rows and the
first two columns.

The complete NNLO prediction for $BR[B \to X_s l^+ l^-]$ depends on
two entries of $\hat{U}^{c(2)}(\mu_b,\mu_0)$, i.e. on
$U^{c(2)}_{72}(\mu_b,\mu_0)$ and $U^{c(2)}_{92}(\mu_b,\mu_0)$ that are
generated by the three-loop matrix $\hat{\gamma}^{c(2)}$.
Unfortunately, the only entries of $\hat{\gamma}^{c(2)}$ that have
been calculated so far are the ones corresponding to the mixing
$\{P_1, ..., P_6\} \to \{P_7,P_8\}$ \cite{CMM97}. Therefore,
$U^{c(2)}_{72}(\mu_b,\mu_0)$ is known but $U^{c(2)}_{92}(\mu_b,\mu_0)$
is not.  Below, we shall include the unknown
$U^{c(2)}_{92}(\mu_b,\mu_0)$ in our analytical formulae. Its potential
numerical relevance will be tested in the next section.

After performing the RGE evolution, one evaluates the perturbative
expression for \linebreak $d\Gamma[b \to X_s l^+ l^-]/d\s$. It amounts
to calculating perturbative matrix elements of the operators $P_i$
among the external partonic on-shell states, multiplying them by the
appropriate Wilson coefficients and performing the phase-space
integrals.  At NLO, one obtains~\cite{M93,BM95}:
\bea 
\f{d \Gamma (b \to X_s l^+ l^-)}{d\s} &=& 
\f{G_{\scs F}^2 m_{b,pole}^5 |V_{ts}^* V_{tb}|^2}{48 \pi^3} 
\left( \f{\alpha_{em}}{4 \pi} \right)^2 (1-\s)^2 
\times \nonumber \\ && \hspace {-3.5cm} \times
\left\{ 
(1 + 2\s) \left( |\tilde{C}_9^{   eff}(\s)|^2 
                    + |\tilde{C}_{10}^{eff}(\s)|^2 \right) 
+ 4 \left( 1 + \f{2}{\s} \right) (\tilde{C}_7^{eff})^2 
+ 12 \tilde{C}_7^{eff} {\rm Re}(\tilde{C}_9^{eff}(\s)) 
\right\}. \hspace{1cm} \label{rate}
\eea
The quantities $\tilde{C}_k^{eff}$ can be split into top- and
light-quark contributions:
\be \label{split}
\tilde{C}_k^{eff}= \tilde{C}_k^{t\;eff} 
+ \f{V_{cs}^* V_{cb}}{V_{ts}^* V_{tb}} \tilde{C}_k^{c\;eff}
+ \f{V_{us}^* V_{ub}}{V_{ts}^* V_{tb}} \left( \tilde{C}_k^{c\;eff}
+ \delta_{k9} \Delta \tilde{C}_9^{eff} \right)
\ee
that are related to the evolved coefficients $C^Q_k(\mu_b)$ as
follows:
\bea
\tilde{C}_7^{Q\;eff} &=& \f{4 \pi}{\al(\mu_b)} C^Q_7(\mu_b)
-\f{1}{3} C^Q_3(\mu_b) -\f{4}{9} C^Q_4(\mu_b)
-\f{20}{3} C^Q_5(\mu_b) -\f{80}{9} C^Q_6(\mu_b),
\label{c7eff.1}\\
\tilde{C}_9^{Q\;eff}(\s) &=& 
4 C^Q_9(\mu_b) \left( \f{\pi}{\al(\mu_b)} + \omega(\s) \right)
+ \sum_{i=1}^6 C^Q_i(\mu_b) \gamma^{Q(0)}_{i9} \ln \f{m_b}{\mu_b}
\nonumber\\ &+& h\left(\f{m_c^2}{m_b^2},\s\right) \left[ \left( 
\f{4}{3} C^c_1(\mu_b) + C^c_2(\mu_b) \right) \delta_{Qc} 
+ 6 C^Q_3(\mu_b) + 60 C^Q_5(\mu_b) \right]
\nonumber\\ &+& h(1,\s) \left(  
-\f{7}{2} C^Q_3(\mu_b)-\f{2}{3} C^Q_4(\mu_b)-38 C^Q_5(\mu_b)-\f{32}{3} C^Q_6(\mu_b) \right)
\nonumber\\ &+& h(0,\s) \left(  
-\f{1}{2} C^Q_3(\mu_b)-\f{2}{3} C^Q_4(\mu_b)- 8 C^Q_5(\mu_b)-\f{32}{3} C^Q_6(\mu_b) \right)
\nonumber\\ &+& 
\f{4}{3} C^Q_3(\mu_b)+ \f{64}{9} C^Q_5(\mu_b)+ \f{64}{27} C^Q_6(\mu_b),
\label{c9eff.1}\\
\tilde{C}_{10}^{Q\;eff}(\s) &=& 
4 C^Q_{10}(\mu_b) \left( \f{\pi}{\al(\mu_b)} + \omega(\s) \right),
\label{c10eff.1}\\
\Delta \tilde{C}_9^{eff} &=& \left[ h(0,\s) - h\left(\f{m_c^2}{m_b^2},\s\right) \right]
                   \left( \f{4}{3} C^c_1(\mu_b) + C^c_2(\mu_b) \right),
\label{dc9eff.1}
\eea
where
\bea
h(z,\s) &=& -\f{4}{9} \ln z +\f{8}{27}+ \f{4}{9}x 
-\f{2}{9}(2+x)\sqrt{|1-x|} \left\{ \begin{array}{ll}
\ln \left|\f{\sqrt{1-x}+1}{\sqrt{1-x}-1}\right|-i\pi, 
                               & {\rm for}\;\;x \equiv 4z/\s < 1,\\
2\;{\rm arctan}(1/\sqrt{x-1}), & {\rm for}\;\;x \equiv 4z/\s > 1,\\
\end{array} \right. \nonumber \\
\omega(\s) &=& -\f{4}{3} Li_2(\s) -\f{2}{3} \ln(1-\s) \ln\s -\f{2}{9} \pi^2
-\f{5+4\s}{3(1+2\s)} \ln(1-\s) \nonumber \\ 
&&-\f{2\s(1+\s)(1-2\s)}{3(1-\s)^2(1+2\s)}\ln \s +\f{5+9\s-6\s^2}{6(1-\s)(1+2\s)}.
\label{h.and.omega}
\eea
Calculating the differential decay rate with the help of eq.~(\ref{rate}),
one {\em must} retain only terms linear in $\omega(\s)$ and also set
$\omega(\s)$ to zero in the interference term proportional to
Re$(C_9^{eff}(\s))$.

The coefficients multiplying $C^Q_1$,..., $C^Q_6$ in
eqs.~(\ref{c7eff.1}) and (\ref{c9eff.1}) are different from the
corresponding ones in refs.~\cite{M93,BM95}, because we use a
different operator basis here.

Substituting the evolved Wilson coefficients to
eqs.~(\ref{c7eff.1})--(\ref{dc9eff.1}), we obtain the following
expressions for the ``effective coefficients'':
\bea
\tilde{C}_7^{c\;eff} &=& - \sum_{i=1}^8 \eta^{a_i} \left[ h^c_i 
+ \f{\al(\mu_0)}{4 \pi} \left( \f{{h'}^{c(-)}_i}{\eta} + {h'}^c_i + {h'}^{cL}_i L \right) \right],
\\[3mm]
\tilde{C}_7^{t\;eff} &=& -\f{1}{2} \eta^{\f{16}{23}} A^t_0(x) 
+\f{4}{3} \left( \eta^{\f{16}{23}} - \eta^{\f{14}{23}} \right) F^t_0(x) 
+ \f{\al(\mu_0)}{4 \pi} \left[ E^t_0(x) \sum_{i=1}^8 {e'}^t_i \eta^{a_i}
\right. \nonumber\\[2mm] && \left.
-\f{1}{2} \eta^{\f{16}{23}} A^t_1(x) 
+\f{4}{3} \left( \eta^{\f{16}{23}} - \eta^{\f{14}{23}} \right) F^t_1(x) 
+ \f{18604}{4761} \left( \eta^{-\f{7}{23}} - \eta^{\f{16}{23}} \right) A^t_0(x) 
\right. \nonumber\\[2mm] && \left.
+\left(  \f{3582208}{357075} \eta^{-\f{9}{23}} 
        -\f{148832}{14283}   \eta^{-\f{7}{23}} 
        -\f{128434}{14283}   \eta^{\f{14}{23}}         
        +\f{3349442}{357075} \eta^{\f{16}{23}} \right) F^t_0(x) \right],
\\[4mm]
\tilde{C}_9^{c\;eff}(\s) &=& -\left( \f{\pi}{\al(\mu_0)} + \f{\omega(\s)}{\eta} \right) 
\sum_{i=3}^9 p^{c(+)}_i \eta^{a_i+1} -\f{1}{4s_w^2} 
\nonumber\\[2mm] && \hspace{-5mm}
-\sum_{i=3}^9 \eta^{a_i} \left[ r^c_i + r^{c(+)}_i \eta + r^{cL(+)}_i \eta L 
+ s^{c}_i \ln \f{m_b}{\mu_b} + t^c_i \;h\left(\f{m_c^2}{m_b^2},\s\right) 
                             + u^c_i h(1,\s) + w^c_i h(0,\s) \right]
\nonumber\\[2mm] && \hspace{-2cm} 
- \f{\al(\mu_0)}{4\pi} \left\{ U^{c(2)}_{92}(\mu_b,\mu_0) + \f{\eta+\omega(\s)}{\eta s_w^2}
+ \sum_{i=3}^9 \eta^{a_i} \left[ {r'}^{c T (+)}_i \eta T(x) +
\f{{r'}^{c(-)}_i}{\eta} + {r'}^c_i + {r'}^{c(+)}_i  \eta  
\right. \right. \nonumber\\[2mm] && \hspace{-1cm}
+ \left( {r'}^{cL}_i + {r'}^{cL(+)}_i \eta \right) L 
+ {r'}^{cL^2(+)}_i \eta L^2 + {r'}^{c \pi^2 (+)}_i \eta \pi^2  
+ \left(  \f{{r'}^{c\omega(-)}_i}{\eta} + {r}^{c(+)}_i + {r}^{cL(+)}_i L \right) 4 \omega(\s)
\nonumber\\[2mm] && 
\;+ \left( \f{{s'}^{c(-)}_i}{\eta}   +   {s'}^c_i   +   {s'}^{cL}_i L \;\right) \ln \f{m_b}{\mu_b} 
\;+ \left( \f{{t'}^{c(-)}_i}{\eta} \;+\; {t'}^c_i \;+\; {t'}^{cL}_i L   \right) h\left(\f{m_c^2}{m_b^2},\s\right) 
\nonumber\\[2mm] && \left. \left.
  + \left( \f{{u'}^{c(-)}_i}{\eta}   +   {u'}^c_i   +   {u'}^{cL}_i L  \right)  h(1,\s) 
  + \left( \f{{w'}^{c(-)}_i}{\eta}   +   {w'}^c_i   +   {w'}^{cL}_i L  \right)  h(0,\s) \right] \right\},
\\[4mm]
\tilde{C}_9^{t\;eff}(\s) &=& \left[ \f{1-4s^2_w}{s^2_w} C^t_0(x)
- \f{1}{s^2_w} B^t_0(x)-D^t_0(x) \right]
\left( 1 + \f{\al(\mu_0)}{\pi} \f{\omega(\s)}{\eta} \right)
\nonumber\\[2mm]  &+& 
\left[ E^t_0(x) + \f{\al(\mu_0)}{4 \pi} \left( E^t_1(x) + \f{4 \omega(\s)}{\eta} E^t_0(x) \right)\right]
\sum_{i=5}^9 q^{t(+)}_i \eta^{a_i+1} 
\nonumber\\[2mm] && \hspace{-2cm} + \f{\al(\mu_0)}{4 \pi} 
\left\{ \f{1-4s^2_w}{s^2_w} C^t_1(x)
-\f{1}{s^2_w} B^t_1(x,-\f{1}{2}) -D^t_1(x) 
+ G^t_1(x) \sum_{i=5}^9 {y'}^{t(+)}_i \eta^{a_i+1}  
\right. \nonumber\\[2mm] && \hspace{-1cm} \left.
+ E^t_0(x) \sum_{i=5}^9 \eta^{a_i} \left[ {r'}^t_i + {r'}^{t(+)}_i \eta + {s'}^t_i \ln \f{m_b}{\mu_b} 
      + {t'}^t_i \;h\left(\f{m_c^2}{m_b^2},\s\right) 
      + {u'}^t_i h(1,\s) + {w'}^t_i h(0,\s) \right] \right\}, \hspace{1cm}
\\[4mm]
\tilde{C}_{10}^{c\;eff}(\s) &=& \f{1}{4s_w^2} 
\left[ 1 + \f{\al(\mu_0)}{\pi}\left(1+\f{\omega(\s)}{\eta} \right) \right],
\\[4mm]
\tilde{C}_{10}^{t\;eff}(\s) &=& \f{1}{s_w^2} 
\left\{ B^t_0(x)-C^t_0(x) + \f{\al(\mu_0)}{4\pi}\left[ B^t_1(x)-C^t_1(x)
+\f{4 \omega(\s)}{\eta} \left( B^t_0(x)-C^t_0(x) \right) \right] \right\}, \hspace{1cm}
\\[4mm]
\Delta \tilde{C}_9^{eff} &=& \left[ h(0,\s) - h\left(\f{m_c^2}{m_b^2},\s\right) \right]
\left\{ -2 \eta^{\f{6}{23}} + \eta^{-\f{12}{23}} 
+ \f{\al(\mu_0)}{4\pi}\left[ 
-\f{15745}{1587} \eta^{-\f{17}{23}} 
\right. \right. \nonumber\\[2mm] && \left. \left. \hspace{3cm}
-\f{ 151}{1587} \eta^{-\f{35}{23}}
-\f{6473}{1587} \eta^{ \f{6}{23}}  
-\f{9371}{1587} \eta^{-\f{12}{23}}
-4L \left( \eta^{\f{6}{23}} + \eta^{-\f{12}{23}} \right) \right] \right\},\\[-7mm] \nonumber
\eea
where $\eta = \al(\mu_0)/\al(\mu_b)$ and $a_i = ( \f{14}{23},
\f{16}{23}, \f{6}{23}, -\f{12}{23}, 0.4086, -0.4230, -0.8994, 0.1456,
-1)_i$.  The ``magic numbers'' entering the above expressions are
collected in tables 1, 2 and 3. 

It is straightforward to verify that our results for the 
${\cal O}(1/\al)$ and ${\cal O}(1)$ parts of $\tilde{C}_9^{eff}$ and
$\tilde{C}_{10}^{eff}$ are identical to the ones found in
refs.~\cite{M93,BM95}. Only the ${\cal O}(\al)$ parts are new here.
As far as $\tilde{C}_7^{eff}$ is concerned, we just reproduce the
result of ref.~\cite{CMM97}, where the ${\cal O}(\al)$ part was already
present.

In order to obtain the complete NLO prediction for the $B \to X_s l^+
l^-$ decay rate, one should \linebreak
\bea
\begin{array}{|l|c|c|c|c|r|r|r|r|}
\hline
~~~i & 1 & 2 & 3 & 4 & 5~~~~ & 6~~~~ & 7~~~~ & 8~~~~ \\ 
\hline
\ &&&&&&&& \\[-4mm]
h^c_i         &  \f{42678}{30253}  & -\f{86697}{103460} &  -\f{3}{7}  &  -\f{1}{14}
              &     \m0.6494       &      \m0.0380      &  \m0.0186   &  \m0.0057 \\[1.5mm]
{h'}^{c(-)}_i \hspace{-1mm}
              &-\f{4246707584}{400095925}
                                   & \f{89606166}{13682585}
                                           & \hspace{-1mm} \f{45043984}{9898119}  \hspace{-1mm} 
                                           & \hspace{-1mm} \f{34505657}{45891279} \hspace{-1mm} 
              &      2.0040        &       0.7476       &  \m0.5385   &    0.0914 \\[1.5mm]
{h'}^c_i      & \hspace{-1mm} \f{3344583818789933}{360615755431797} \hspace{-1mm} 
                           & \hspace{-1.5mm} -\f{90790555261878016}{13088650734603675} \hspace{-1mm} 
                                                        &  -\f{6473}{7406}   
                                                                      &  \f{9371}{22218} 
              &     \m2.7231       &       0.4083       &    0.1465   &    0.0205 \\[1.5mm]
{h'}^{cL}_i   & \f{199164}{30253}  &-\f{115596}{25865}  & -\f{6}{7}   &  \f{2}{7} 
              &     \m2.0343       &       0.1232       &    0.1279   &  \m0.0064 \\[1.5mm]
{e'}^t_i      &\f{4298158}{816831} &  -\f{8516}{2217}   &      0      &      0
              &     \m1.9043       &      \m0.1008      &      0.1216 &    0.0183 \\[1.5mm]
\hline
\end{array} \nonumber
\eea
\begin{center}
Table 1. ``Magic numbers'' entering the expressions for 
           $\tilde{C}_7^{c\;eff}$ and $\tilde{C}_7^{t\;eff}$.
           Three-loop anomalous dimensions from ref.~\cite{CMM97}
           have been used in their evaluation.
\end{center}

\newpage
\bea
\begin{array}{|l|c|c|r|r|r|r|c|}
\hline
~~~i & 3 & 4 & 5~~~~ & 6~~~~ & 7~~~~ & 8~~~~ & 9 \\ 
\hline
\ &&&&&&& \\[-4mm]
p^{c(+)}_i & -\f{80}{203} &  \f{8}{33} &  
           0.0433 &  0.1384 &  0.1648  & -0.0073 & -\f{4704688}{25088393} \\[1.5mm]
r^c_i & \f{3085}{3703} & -\f{129}{1058} &     
          -0.1642 &  0.0793 & -0.0451  & -0.1638 & 0 \\[1.5mm]
r^{c(+)}_i & -\f{64730}{322161} & -\f{18742}{52371} &
           0.0454 & -0.3719 & -0.3254 &  0.0066 & \f{1775737}{809303} \\[1.5mm]
r^{cL(+)}_i & -\f{40}{203} & -\f{8}{33} &  
           0.0339 & -0.1122 & -0.2841 & -0.0020 & \f{27051956}{75265179} \\[1.5mm]
{r'}^{c T (+)}_i & \f{20}{609} & \f{4}{99} & 
          -0.0021 &  0.0289 &  0.0174 &  0.0010 & -\f{8908520}{75265179} \\[1.5mm]
{r'}^{c(-)}_i & -\f{316900}{299943} & \f{51388}{128547} & 
           1.9957 & -0.8153 &  0.1488 & -0.2353 & 0 \\[1.5mm]
{r'}^c_i & \f{183859}{42849} & \f{130739}{128547} & 
          -0.0939 & -0.9763 &  0.0393 & -2.2799 & 0 \\[1.5mm]
{r'}^{c(+)}_i & -\f{8129495}{5798898} & -\f{4447705}{942678} & 
           0.6261 & -3.6869 &  0.2246 &  0.0121 & \f{4896690443}{677386611} \\[1.5mm]
{r'}^{cL}_i & \f{6170}{3703} & \f{258}{529} & 
          -0.5145 & -0.2571 &  0.3111 & -0.1829 & 0 \\[1.5mm]
{r'}^{cL(+)}_i & -\f{97850}{33327} & -\f{398258}{157113} & 
           0.6618 & -2.2108 & -1.6839 &  0.0472 & \f{4704688}{2595351} \\[1.5mm]
{r'}^{cL^2(+)}_i & -\f{20}{21} & -\f{4}{9} & 
           0.1833 & -0.2481 & -0.1096 & -0.0090 & 0 \\[1.5mm]
{r'}^{c\pi^2(+)}_i & -\f{20}{63} & -\f{4}{27} & 
           0.0611 & -0.0827 & -0.0365 & -0.0030 & 0 \\[1.5mm]
{r'}^{c\omega(-)}_i & \f{87527}{99981} & -\f{6217}{85698} & 
          -0.1685 &  0.0323 & -0.0475 & -0.2018 & 0 \\[1.5mm]
s^c_i & -\f{40}{21} & \f{4}{9} & 
           0.2340 &  0.3061 &  0.0636 & -0.0322 & 0 \\[1.5mm]
{s'}^{c(-)}_i & -\f{1373012}{128547} & -\f{735748}{385641} &
           2.1605 &  0.3356 &  0.8434 & -0.2456 & 0 \\[1.5mm]
{s'}^c_i & -\f{129460}{33327} & -\f{37484}{14283} &
           0.9813 & -3.2900 & -0.5020 &  0.1151 & 0 \\[1.5mm]
{s'}^{cL}_i & -\f{80}{21} & -\f{16}{9} &
           0.7330 & -0.9925 & -0.4383 & -0.0359 & 0 \\[1.5mm]
t^c_i & \f{12}{7} & -\f{2}{3} &      
           0.1658 & -0.2407 & -0.0717 &  0.0990 & 0 \\[1.5mm]
{t'}^{c(-)}_i & \f{33606}{3703} & -\f{6046}{4761} & 
          -0.1681 &  1.2986 & -0.3397 &  0.4766 & 0 \\[1.5mm]
{t'}^c_i & \f{12946}{3703} & \f{18742}{4761} & 
           0.6951 &  2.5871 &  0.5664 & -0.3540 & 0 \\[1.5mm]
{t'}^{cL}_i & \f{24}{7} & \f{8}{3} & 
           0.5193 &  0.7805 &  0.4945 &  0.1106 & 0 \\[1.5mm]
u^c_i    & \f{2}{7} & 0 &     
          -0.2559 &  0.0083 &  0.0180 & -0.0562 & 0 \\[1.5mm]
{u'}^{c(-)}_i & \f{168155}{99981} & \f{166}{81} & 
          -1.0892 & -1.1627 & -0.2197 & -0.2193 & 0 \\[1.5mm]
{u'}^c_i & \f{6473}{11109} & 0 & 
          -1.0733 & -0.0897 & -0.1424 &  0.2008 & 0 \\[1.5mm]
{u'}^{cL}_i & \f{4}{7} & 0 & 
          -0.8018 & -0.0271 & -0.1243 & -0.0627 & 0 \\[1.5mm]
w^c_i & \f{1}{7} & \f{1}{6} &
          -0.1731 & -0.1120 & -0.0178 & -0.0067 & 0 \\[1.5mm]
{w'}^{c(-)}_i & \f{251737}{199962} & \f{117137}{85698} & 
          -1.1732 & -0.5134 & -0.3895 &  0.0190 & 0 \\[1.5mm]
{w'}^c_i & \f{6473}{22218} & -\f{9371}{9522} & 
          -0.7257 &  1.2038 &  0.1408 &  0.0238 & 0 \\[1.5mm]
{w'}^{cL}_i & \f{2}{7} & -\f{2}{3} & 
          -0.5421 &  0.3632 &  0.1229 & -0.0074 & 0 \\[1.5mm]
\hline
\end{array} \nonumber 
\eea
\ \vspace{-1cm}
\begin{center}
Table 2. ``Magic numbers'' entering the expression for $\tilde{C}_9^{c\;eff}$.
\end{center}

\newpage
\bea
\begin{array}{|l|r|r|r|r|c|}
\hline
~~~i & 5~~~~ & 6~~~~ & 7~~~~ & 8~~~~ & 9 \\ 
\hline
\ &&&&& \\[-4mm]
q^{t(+)}_i    &  0.0318 &  0.0918 & -0.2700 &  0.0059 & \f{33160}{235941}     \\[1.5mm]
{r'}^t_i      & -0.4817 &  0.2104 &  0.2956 &  0.5246 & 0                     \\[1.5mm]
{r'}^{t(+)}_i &  0.2164 & -0.4330 & -0.9126 &  0.0660 & \f{6672596}{12976755} \\[1.5mm]
{s'}^t_i      &  0.6862 &  0.8125 & -0.4165 &  0.1031 & 0                     \\[1.5mm]
{t'}^t_i      &  0.4861 & -0.6389 &  0.4699 & -0.3171 & 0                     \\[1.5mm]
{u'}^t_i      & -0.7505 &  0.0221 & -0.1182 &  0.1799 & 0                     \\[1.5mm]
{w'}^t_i      & -0.5075 & -0.2973 &  0.1168 &  0.0213 & 0                     \\[1.5mm]
{y'}^{t(+)}_i & -0.1242 & -0.0956 & -0.1628 & -0.0176 & \f{157366}{393235}    \\[1.5mm]
\hline
\end{array} \nonumber
\eea
\begin{center}
Table 3. ``Magic numbers'' entering the expression for $\tilde{C}_9^{t\;eff}$.
\end{center}
use eqs.~(\ref{rate})--(\ref{dc9eff.1}) and neglect the ${\cal
  O}(\al)$ contributions to the effective coefficients
$\tilde{C}_k^{Q\;eff}(\s)$ (i.e. include only the ${\cal O}(1/\al)$
and ${\cal O}(1)$ parts of them). On the other hand, in the complete
NNLO calculation, it is {\em not} sufficient to take into account the
${\cal O}(\al)$ parts of the effective coefficients.  One should also
modify eq.~(\ref{rate}) by including effects originating e.g. from
two-loop matrix elements of the four-quark operators and the
corresponding Bremsstrahlung corrections.

In the present paper, we are able to include the NNLO effects only
partly.  We shall simply use eq.~(\ref{rate}), but at the same time we
will include the ${\cal O}(\al)$ contributions to the effective
coefficients. In this way, we will include all the $m_t$-dependent
NNLO contributions to the branching ratio,\footnote{
  The only exceptions are the $m_t$-dependent contributions from the
  one-loop matrix elements of $P_7$ and $P_8$. However, they are
  proportional to the relatively small Wilson coefficients
  $C_7(\mu_b)$ and $C_8(\mu_b)$ that do not grow with $m_t$ in the
  formal limit $m_t \to \infty$.}
as well as the terms enhanced by $1/s^2_w \sim 4.3$. It is
important to calculate the $m_t$-dependent terms at the NNLO level,
because both $C^t_9(\mu_0)$ and $C^t_{10}(\mu_0)$ grow with $m_t$ in
the formal limit $m_t \to \infty$.  Therefore, $m_t^2/M_W^2 \sim 4.8$
plays the role of an enhancement factor, too.

Above, we have presented explicitly all the ${\cal O}(\al)$ parts of
the effective coefficients. However, the unknown quantity
$U^{c(2)}_{92}(\mu_b,\mu_0)$ occurred in $\tilde{C}_9^{c\;eff}(\s)$.
In our numerical calculations described in the next section, it will
be assumed that $U^{c(2)}_{92}(\mu_b,\mu_0)$ vanishes. We shall relax
this assumption below eq.~(\ref{numR}), and check that the expected
numerical effect of $U^{c(2)}_{92}(\mu_b,\mu_0)$ on the decay rate is
very small. \vspace{1cm}

\section{Phenomenological implications}
\label{phenom}

In the present section, we shall study the numerical importance of the
calculated NNLO effects as well as the uncertainties due to the yet
unknown contributions.

As a first step, let us calculate the effective coefficients for
several different values of $\mu_0$ and $\mu_b$. We will vary $\mu_b$
by a factor of 2 around $m_b \sim 5$~GeV, i.e. we will take
$\mu_b =$~2.5,~5~and~10~GeV. In the expressions for $\tilde{C}_k^{c\;eff}$
and $\Delta \tilde{C}_9^{eff}$,
we will vary $\mu_0$
by a factor of 2 around $M_W \sim 80$~GeV, i.e. we will take
$\mu_0 =$~40,~80~and~160~GeV. In the expressions for $\tilde{C}_k^{t\;eff}$,
we will vary $\mu_0$
by a factor of 2 around $\sqrt{M_W m_t} \sim 120$~GeV, i.e. we will take
$\mu_0 =$~60,~120~and~240~GeV. 

The remaining input parameters will be equal to \cite{PData98}
\bea
\al(M_Z) = 0.119,                 \hspace{1cm}
m_t^{pole} = 173.8~{\rm GeV},     \hspace{1cm}
M_W = 80.41~{\rm GeV},          \hspace{1cm}
s^2_w = 0.23124.
\nonumber
\eea
Since we shall keep $\s$ arbitrary, our expressions for
$\tilde{C}_9^{Q\;eff}$, $\tilde{C}_{10}^{Q\;eff}$ and $\Delta
\tilde{C}_9^{eff}$ will read
\bea
\tilde{C}_9^{   Q\;eff} &=& A_9^Q    + R_9^Q       \omega(\s) 
     + T_9^Q \;h\left(\f{m_c^2}{m_b^2},\s\right) + U_9^Q h(1,\s) + W_9^Q h(0,\s),\\
\tilde{C}_{10}^{Q\;eff} &=& A_{10}^Q + R_{10}^Q \; \omega(\s),\\
\Delta \tilde{C}_9^{eff} &=& Z_9 \left[ h(0,\s) - h\left(\f{m_c^2}{m_b^2},\s\right) \right].
\eea
The coefficients $A_k^Q$, ..., $W_k^Q$ are independent of $m_c$, and
they only weakly depend on $m_b$ via the logarithm $\ln(m_b/\mu_b)$.
In this logarithm, we shall use $m_b = 4.8$~GeV.

In tables 4 and 5, our results for $\tilde{C}_7^{Q\;eff}$,
$A_k^Q$, ..., $W_k^Q$ and $Z_9$ are given, both with and without the
${\cal O}(\al)$ contributions. They allow the following observations:
\begin{itemize}
\item{} The dominant contributions to the ``effective coefficients''
  and to the decay rate originate from $A^c_9$ and $A^t_{10}$.
  However, the coefficients $\tilde{C}_7^{Q\;eff}$ are not much less
  important, because of the factor ``12'' in the last term of
  eq.~(\ref{rate}). 
\item{} The inclusion of the ${\cal O}(\al)$ contributions
  significantly reduces the $\mu_0$-dependence. It is especially
  important in the case of $A^t_{10}$, which had varied by more than
  $\pm$10\% before including the ${\cal O}(\al)$ correction. The
  dependence on $\mu_0$ remains significant only in the relatively
  small quantities such as $R_{10}^t$. ($R_{10}^t$ is multiplied by
  $\omega(\s) \in [-1.32,-1.24]$ for $\s \in [0.05,0.25]$).
\end{itemize}

\begin{center}
\begin{tabular}{|l|c|c|c||c|c|}
\hline
$\mu_0$ [GeV] & 40     & 80     & 160     &  80     & 80     \\
$\mu_b$ [GeV] &  5     &  5     &   5     &   2.5   & 10     \\
\hline \hline
$\al(\mu_0)$  &  0.136 &  0.121 &   0.110 &   0.121 &  0.121 \\
$\al(\mu_b)$  &  0.215 &  0.215 &   0.215 &   0.267 &  0.180 \\
$\eta$        &  0.633 &  0.565 &   0.510 &   0.454 &  0.674 \\
\hline
$\tilde{C}_7^{c\;eff}$ with ${\cal O}(\al)$ 
& 0.567 & 0.567 & 0.566 & 0.554 & 0.579 \\
$\tilde{C}_7^{c\;eff}$ without ${\cal O}(\al)$ 
& 0.631 & 0.631 & 0.632 & 0.634 & 0.631 \\
\hline
$A_9^c$ with ${\cal O}(\al)$ 
& $-$4.685 & $-$4.683 & $-$4.689 & $-$4.612 & $-$4.828 \\
$A_9^c$ without ${\cal O}(\al)$ 
& $-$4.620 & $-$4.635 & $-$4.681 & $-$4.750 & $-$4.635 \\
$A_9^c$ only ${\cal O}(1/\al)$ 
& $-$1.569 & $-$1.964 & $-$2.315 & $-$2.181 & $-$1.612 \\
\hline
$R_9^c$ with ${\cal O}(\al)$ 
& $-$0.315 & $-$0.316 & $-$0.320 & $-$0.415 & $-$0.242 \\
$R_9^c$ without ${\cal O}(\al)$ 
& $-$0.107 & $-$0.134 & $-$0.158 & $-$0.186 & $-$0.092 \\
\hline
$T_9^c$ with ${\cal O}(\al)$ 
& $-$0.641 & $-$0.625 & $-$0.603 & $-$0.393 & $-$0.807 \\
$T_9^c$ without ${\cal O}(\al)$ 
& $-$0.505 & $-$0.374 & $-$0.255 & $-$0.115 & $-$0.576 \\
\hline
$U_9^c$ with ${\cal O}(\al)$ 
& $-$0.048 & $-$0.050 & $-$0.052 & $-$0.070 & $-$0.035 \\
$U_9^c$ without ${\cal O}(\al)$ 
& $-$0.026 & $-$0.032 & $-$0.038 & $-$0.045 & $-$0.022 \\
\hline
$W_9^c$ with ${\cal O}(\al)$ 
& $-$0.045 & $-$0.046 & $-$0.047 & $-$0.062 & $-$0.033 \\
$W_9^c$ without ${\cal O}(\al)$ 
& $-$0.026 & $-$0.032 & $-$0.038 & $-$0.044 & $-$0.022 \\
\hline
$A_{10}^c$ with ${\cal O}(\al)$ 
& 1.128 & 1.123 & 1.119 & 1.123 & 1.123 \\
$A_{10}^c$ without ${\cal O}(\al)$ 
& 1.081 & 1.081 & 1.081 & 1.081 & 1.081 \\
\hline
$R_{10}^c$ with ${\cal O}(\al)$ 
& 0.074 & 0.074 & 0.074 & 0.092 & 0.062 \\
$R_{10}^c$ without ${\cal O}(\al)$ 
& 0 & 0 & 0 & 0 & 0 \\
\hline
$Z_9$ with ${\cal O}(\al)$ 
& $-$0.648 & $-$0.634 & $-$0.613 & $-$0.410 & $-$0.811 \\
$Z_9$ without ${\cal O}(\al)$ 
& $-$0.506 & $-$0.376 & $-$0.257 & $-$0.118 & $-$0.577 \\
\hline
\end{tabular}\\[2mm]
Table 4. $\tilde{C}_7^{c\;eff}$, $A_k^c$, ..., $W_k^c$ and $Z_9$ for
various values of $\mu_0$ and $\mu_b$.
\end{center}
\begin{itemize}
\item{} The dependence on $\mu_b$ remains rather strong in most of the
  listed quantities. It follows mainly from the fact that two-loop
  matrix elements of the four-quark operators have not been included.
  It is relevant especially to the cases of $C_7^{t\;eff}$, $T^c_9$
  and $R^c_9$, which will cause considerable $\mu_b$-dependence of the
  final prediction for the decay rate.
\item{} The coefficients $W^c_9$ turn out to be very small, while
  $\Delta \tilde{C}_9^{eff}$ in eq.~(\ref{split}) is multiplied by
  $|(V_{us}^* V_{ub})/(V_{ts}^* V_{tb})| \simeq 0.08$.  In
  consequence, the terms containing $h(0,\s)$ contribute by less than
  3\% to the differential decay rate for $\s > 0.05$, because
  $|h(0,\s)| = |\f{8}{27} - \f{4}{9}(\ln\s-i\pi)|\;$ is smaller than
  2.2 in this region.  This is fortunate, because $h(0,\s)$ is
  expected to receive huge non-perturbative contributions from
  intermediate light hadron states.\footnote{
These contributions are expected to be of the same size as $h(0,\s)$ itself,
after taking an average over a sufficiently wide region of $\s$.}
The smallness of $W^c_9$ and $V_{ub}$ allows us to use only the
perturbative expression for $h(0,\s)$ below. We could equivalently
just neglect it.
\end{itemize}
\begin{center}
\begin{tabular}{|l|c|c|c||c|c|}
\hline
$\mu_0$ [GeV] &  60     & 120    & 240     & 120     & 120     \\
$\mu_b$ [GeV] &   5     &  5     &   5     &   2.5   & 10     \\
\hline \hline
$m_t^{\overline{MS}}(\mu_0)$ [GeV]
              & 180     & 170    & 162     & 170     & 170 \\
$\al(\mu_0)$  &   0.127 &  0.114 &   0.104 &   0.114 &   0.114 \\ 
$\al(\mu_b)$  &   0.215 &  0.215 &   0.215 &   0.267 &   0.180 \\
$\eta$        &   0.591 &  0.531 &   0.483 &   0.427 &   0.635 \\
\hline
$\tilde{C}_7^{t\;eff}$ with ${\cal O}(\al)$ 
& 0.261 & 0.265 & 0.266 & 0.225 & 0.300 \\
$\tilde{C}_7^{t\;eff}$ without ${\cal O}(\al)$ 
& 0.325 & 0.310 & 0.297 & 0.274 & 0.344 \\
\hline
$A_9^t$ with ${\cal O}(\al)$ 
& $-$0.547 & $-$0.541 & $-$0.544 & $-$0.541 & $-$0.542 \\
$A_9^t$ without ${\cal O}(\al)$ 
& $-$0.425 & $-$0.506 & $-$0.579 & $-$0.509 & $-$0.504 \\
\hline
$R_9^t$ with ${\cal O}(\al)$ 
& $-$0.029 & $-$0.035 & $-$0.040 & $-$0.043 & $-$0.029 \\
$R_9^t$ without ${\cal O}(\al)$ 
& 0 & 0 & 0 & 0 & 0\\
\hline
$T_9^t$ with ${\cal O}(\al)$ 
& 0.0002 & 0.0003 & 0.0004 & 0.0005 & 0.0001 \\
$T_9^t$ without ${\cal O}(\al)$ 
& 0 & 0 & 0 & 0 & 0\\
\hline
$U_9^t$ with ${\cal O}(\al)$ 
& $-$0.002 & $-$0.002 & $-$0.002 & $-$0.002 & $-$0.002 \\
$U_9^t$ without ${\cal O}(\al)$ 
& 0 & 0 & 0 & 0 & 0\\
\hline
$W_9^t$ with ${\cal O}(\al)$ 
& $-$0.002 & $-$0.002 & $-$0.002 & $-$0.002 & $-$0.002 \\
$W_9^t$ without ${\cal O}(\al)$ 
& 0 & 0 & 0 & 0 & 0\\
\hline
$A_{10}^t$ with ${\cal O}(\al)$ 
& $-$3.051 & $-$3.115 & $-$3.107 & $-$3.115 & $-$3.115 \\
$A_{10}^t$ without ${\cal O}(\al)$ 
& $-$3.688 & $-$3.292 & $-$2.964 & $-$3.292 & $-$3.292 \\
\hline
$R_{10}^t$ with ${\cal O}(\al)$ 
& $-$0.252 & $-$0.225 & $-$0.203 & $-$0.280 & $-$0.189 \\
$R_{10}^t$ without ${\cal O}(\al)$ 
& 0 & 0 & 0 & 0 & 0 \\
\hline
\end{tabular}\\[2mm]
Table 5. $\tilde{C}_7^{t\;eff}$ and $A_k^t$, ..., $W_k^t$ for various
values of $\mu_0$ and $\mu_b$. 
\end{center}

\begin{figure}
\vspace{-2.5cm} 
\includegraphics[width=7.5cm,angle=0]{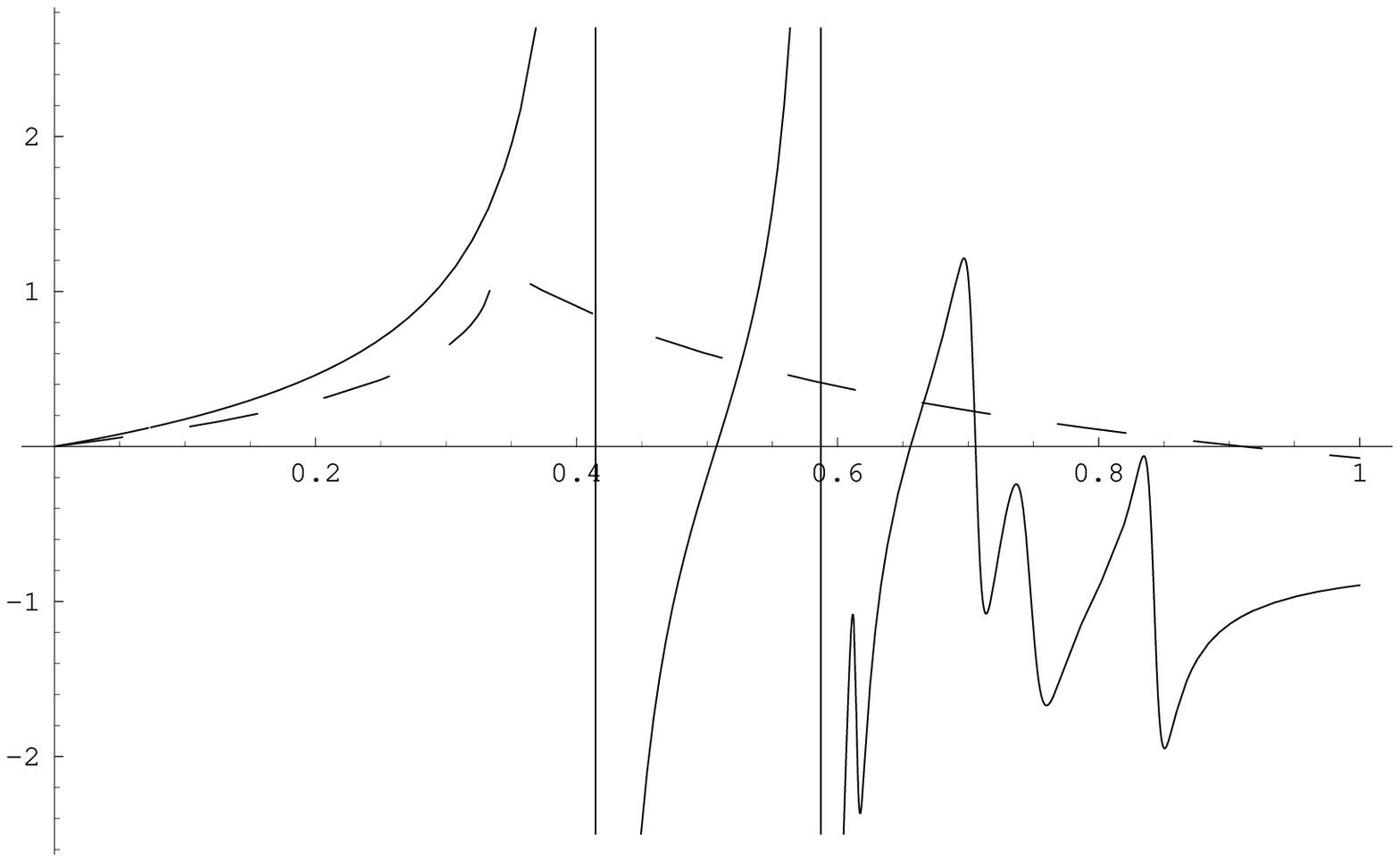}
\hspace{1cm}
\includegraphics[width=7.5cm,angle=0]{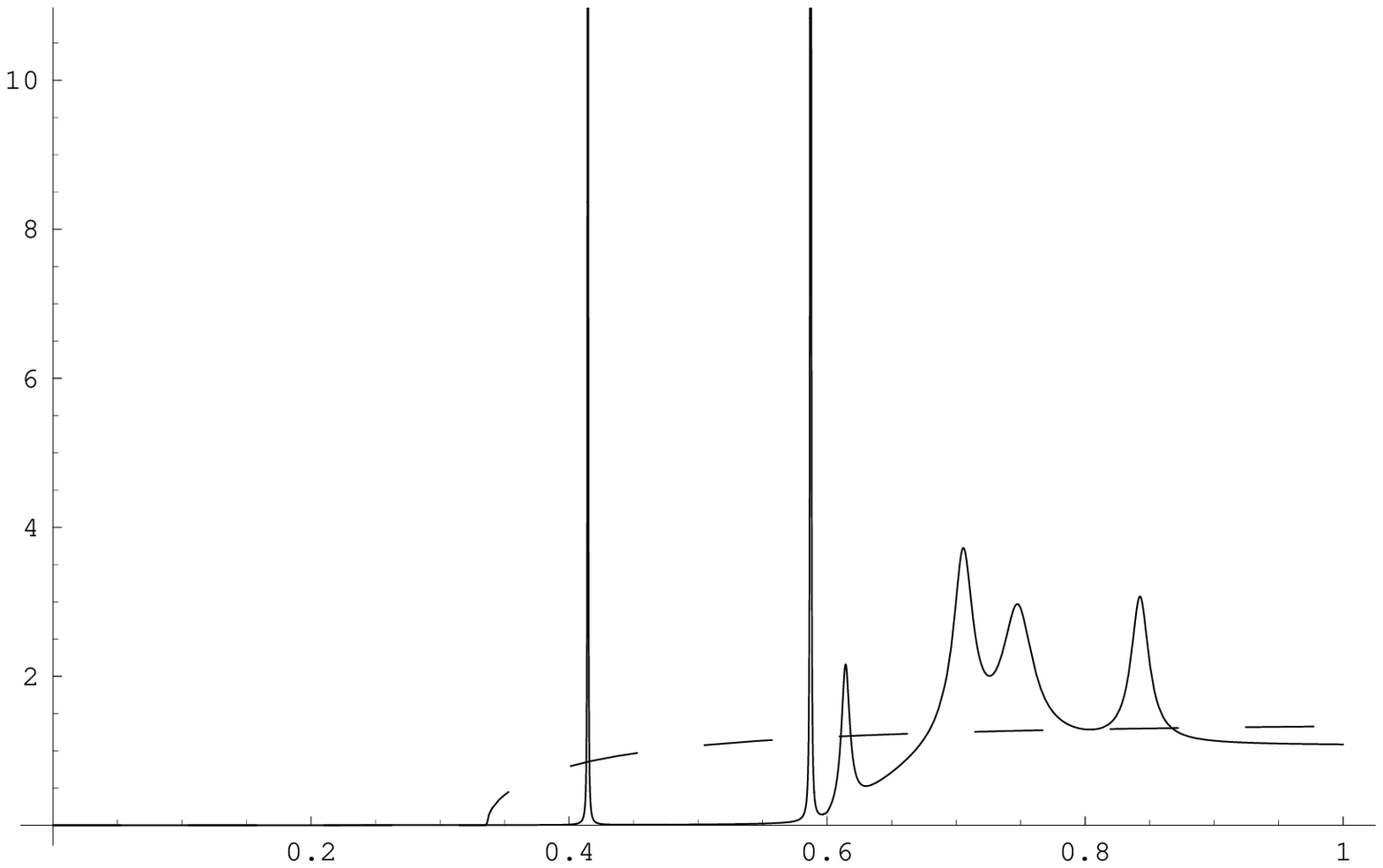}
\vspace{-3.2cm}
\begin{center}
\caption{Perturbative and non-perturbative versions of
Re$[h(m_c^2/m_b^2,\s)-h(m_c^2/m_b^2,0)]$ 
and
Im$[h(m_c^2/m_b^2,\s)-h(m_c^2/m_b^2,0)]$ 
as functions of $\s$ (see the text).}
\label{hzs}
\end{center}
\vspace{-1cm}
\end{figure}

Huge non-perturbative contributions occur in $h(m_c^2/m_b^2,\s)$ as
well, for $\s > (2m_c/m_b)^2$. It is illustrated in fig.~\ref{hzs}.
Dashed lines show the real and imaginary parts of $h(z,\s)$ from
eq.~(\ref{h.and.omega}), with $z=(1.4/4.8)^2$ and with $h(z,0)$
subtracted. Solid lines present non-perturbative estimates of the same
quantities obtained using the formulae and parameters from
ref.~\cite{KS96} where the factorization approximation and dispersion
relations were used.\footnote{
  However $4 m_D^2$ is replaced by $4 m_{\pi}^2$ in eq.~(3.4) of
  ref.~\cite{KS96}.  We thank F.~Kr\"uger for confirming that this was
  a misprint.}

While the solid lines in fig.~\ref{hzs} should not be regarded as the
true non-perturbative results (because of the factorization
approximation), they give us qualitative information on the size of
expected non-perturbative effects. In particular, we can observe that
replacing the solid lines by the dashed ones in the region $\s \in
[0.05,0.25]$ should have quite a small effect on the predicted
differential decay rate, owing to the relatively small size of $T^Q_9$
in tables 4 and 5. Actually, the $\mu_b$-dependence of $T^c_9$ is
numerically more important. Our aim below will be predicting the decay
rate integrated over $\s$ from 0.05 to 0.25. We shall use the purely
perturbative expression for $h(z,\s)$, keeping in mind that the
$\mu_b$-dependence of our prediction is expected to be larger than the
uncertainty stemming from neglected non-perturbative
effects.\footnote{
The non-perturbative effects estimated in fig.~\ref{hzs} are not included
in the HQET correction we shall take into account later.}

As far as $h(1,\s)$ is concerned, the argument for using the purely
perturbative expression can be the same as for $h(0,\s)$ (small
coefficients) or the same as for $h(m_c^2/m_b^2,\s)$ (convergence of
the perturbative and non-perturbative results for small $\s$).\\

The decay rate given in eq.~(\ref{rate}) suffers from large
uncertainties due to $m_{b,pole}^5$ and the CKM angles. One can get
rid of them by normalizing to the semileptonic decay rate of the
$b$-quark
\be
\Gamma[ b \to X_c e \bar{\nu}_e] = 
\f{ G_F^2 m_{b,pole}^5 }{ 192 \pi^3} |V_{cb}|^2
g\left(\f{m_{c,pole}^2}{m_{b,pole}^2} \right) 
\kappa\left(\f{m_c^2}{m_b^2} \right),
\ee
where 
\be \label{g(z)}
g(z) = 1 - 8 z + 8 z^3 - z^4 - 12 z^2 \ln z 
\ee
is the phase-space factor, and 
\be  \label{kappa}
\kappa(z) = 1 - \f{2 \al (m_b)}{3 \pi} \f{h(z)}{g(z)}
\ee
is a sizeable next-to-leading order QCD correction to the semileptonic
decay \cite{CM78}. The function $h(z)$ has been given analytically in
ref.~\cite{N89}:
\bea
h(z) =
- (1-z^2) \left( \f{25}{4}- \f{239}{3} z + \f{25}{4} z^2 \right) 
+ z \ln z \left( 20 + 90 z - \f{4}{3} z^2 + \f{17}{3} z^3 \right) 
+ z^2 \ln^2 z \; ( 36 + z^2) 
\hspace{1.5cm} && \nonumber \\
+ (1-z^2) \left( \f{17}{3} - \f{64}{3} z + \f{17}{3} z^2 \right) \ln (1-z) 
- 4 ( 1 + 30 z^2 + z^4 ) \ln z \; \ln (1-z) 
\hspace{5cm} && \nonumber \\
- (1 + 16z^2 + z^4) [ 6 {\rm Li}_2(z) - \pi^2 ] 
- 32 z^{3/2} (1+z) \left[ \pi^2 
     - 4 {\rm Li}_2(\sqrt{z}) +  4 {\rm Li}_2(-\sqrt{z}) 
     - 2 \ln z \; \ln \left( \f{1-\sqrt{z}}{1+\sqrt{z}} \right) \right]. 
\hspace{0.5cm} && \nonumber \eea 

        Thus, the final perturbative quantity we consider is the ratio
\be \label{ratio}
R^{l^+l^-}_{quark}(\s) = \f{1}{\Gamma[ b \to X_c e \bar{\nu}_e ]}
\f{d}{d\s} \Gamma (b \to X_s l^+ l^-). 
\ee

\begin{figure}[h]
\hspace{ 4mm} $R^{l^+l^-}_{quark}(\s)~[10^{-4}]$ 
\hspace{51mm} $R^{l^+l^-}_{quark}(\s)~[10^{-4}]$ \\[-38mm] 
\includegraphics[width=8cm,angle=0]{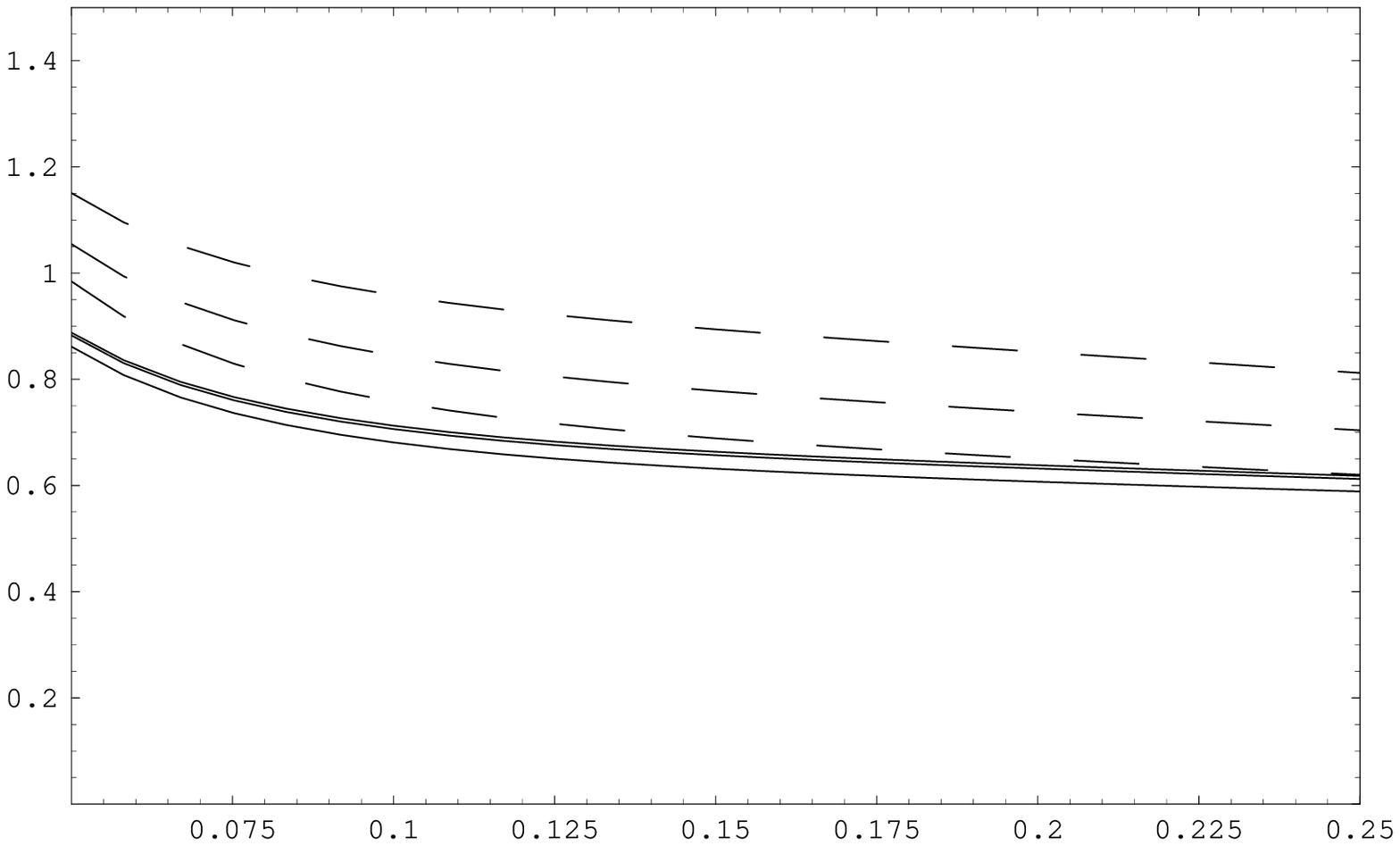}
\includegraphics[width=8cm,angle=0]{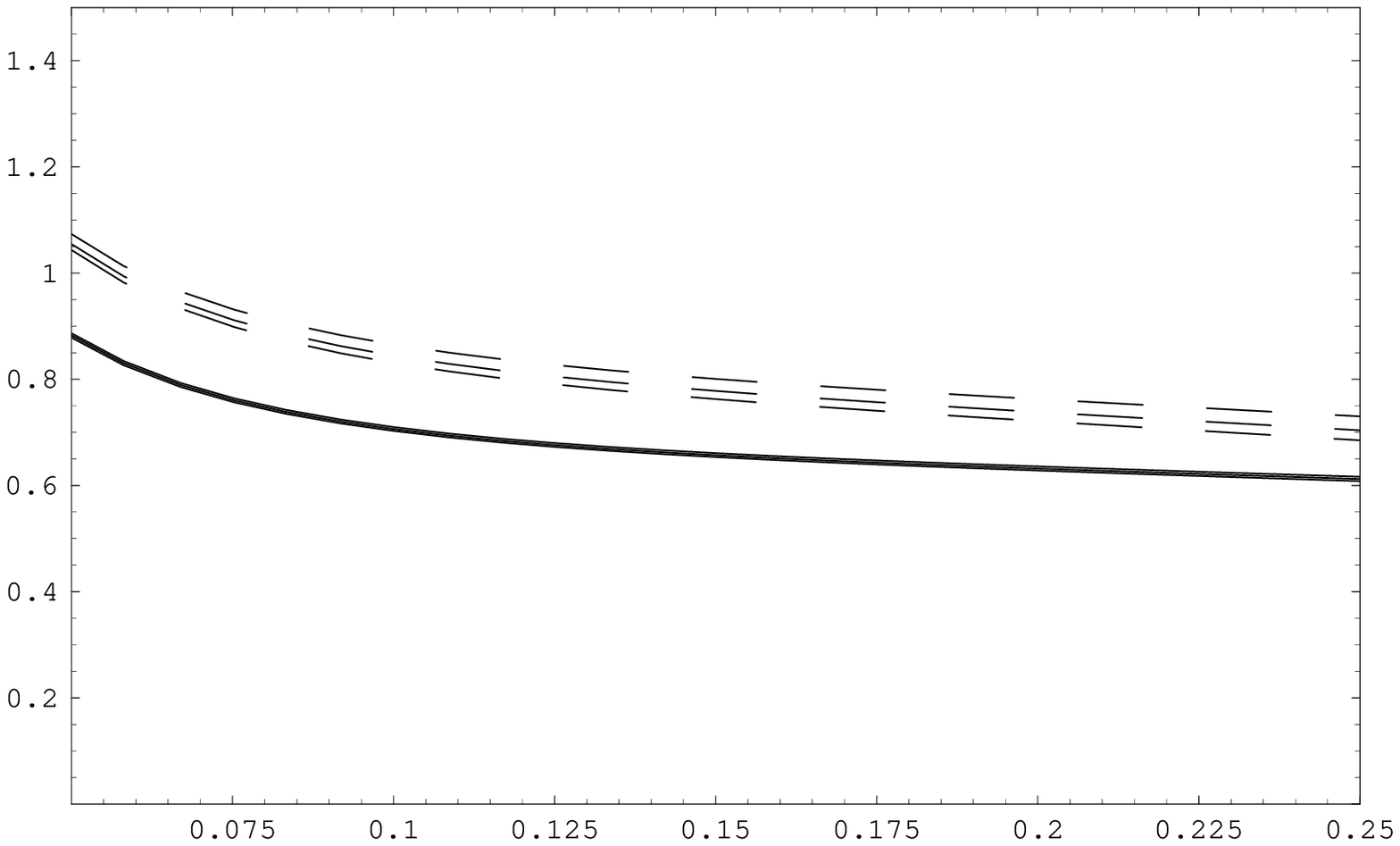} \\[-41mm]
\hspace*{39mm} $\s$ \hspace{76mm} $\s$
\begin{center}
\caption{Reduction of $\mu_0$-dependence of $R^{l^+l^-}_{quark}(\s)$.}
\label{mu0dep}
\end{center}
\vspace{-5mm}
\end{figure}
\begin{figure}[h]
\hspace{46mm} $R^{l^+l^-}_{quark}(\s)~[10^{-4}]$ \\[-45mm] 
\begin{center}
\includegraphics[width=8cm,angle=0]{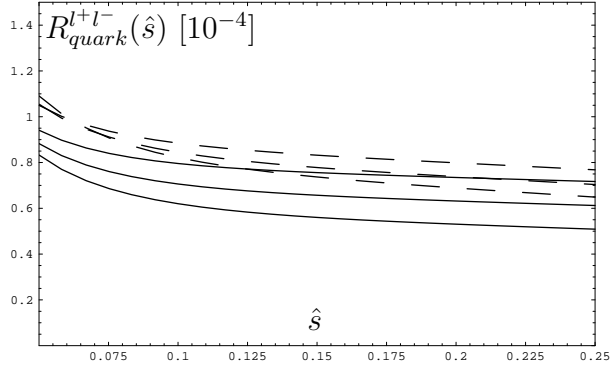}\\[-41mm]
~$\s$\\[3mm]
\caption{Remaining $\mu_b$-dependence of $R^{l^+l^-}_{quark}(\s)$.}
\label{mubdep}
\end{center}
\vspace{-5mm}
\end{figure}

Our results for $R^{l^+l^-}_{quark}(\s)$ in the domain $\s \in
[0.05,0.25]$ are presented in figs.~\ref{mu0dep} and \ref{mubdep}.  In
their evaluation, we have used $\alpha_{em} = \alpha_{em}(m_b
\sqrt{0.15}) = \f{1}{133}$  and $|V_{ts}^* V_{tb}/V_{cb}| = 0.976$.
The quantity $\Delta \tilde{C}_9^{eff}$ that is multiplied by $V_{ub}$
has been neglected. The dashed lines represent the pure NLO results,
i.e. the ones with neglected ${\cal O}(\al)$ parts of the effective
coefficients. The solid lines are obtained after including the 
${\cal O}(\al)$ terms. Some of them overlap, and look like thick
lines.

In both plots of fig.~\ref{mu0dep}, $\mu_b=5$~GeV, and three different
values of $\mu_0$ are chosen.  The left plot corresponds to varying
$\mu_0$ by a factor of 2 around $\sqrt{M_W m_t}$ in
$\tilde{C}_k^{t\;eff}$ (as in the first three columns of table 5) and
keeping it fixed to $M_W$ in $\tilde{C}_k^{c\;eff}$.  The right plot
corresponds to varying $\mu_0$ by a factor of 2 around $M_W$ in
$\tilde{C}_k^{c\;eff}$ (as in the first three columns of table 4) and
keeping it fixed to $\sqrt{M_W m_t}$ in $\tilde{C}_k^{t\;eff}$.

The importance of including the two-loop matching conditions is
clearly seen: the dependence on $\mu_0$ decreases from $\pm16\%$ to
around $\pm2.5\%$ at the representative point $\s=0.2$.  Most of the
effect is due to the strong $m_t$-dependence of $A^t_{10}$ and to the
$\mu_0$-dependence of $m_t^{\overline{MS}}(\mu_0)$.

In fig.~\ref{mubdep}, the scale $\mu_0$ is fixed to 120~GeV in
$\tilde{C}_k^{t\;eff}$ and to 80~GeV in $\tilde{C}_k^{c\;eff}$, while
the scale $\mu_b$ takes the values of 2.5, 5 and 10~GeV. One can see
that the $\mu_b$-dependence increases after taking into account the
${\cal O}(\al)$ contributions to the effective coefficients.  When the
${\cal O}(\al)$ terms are not included, an accidental cancellation of
the $\mu_b$-dependence occurs among the four contributions to the
differential decay rate in eq.~(\ref{rate}). This cancellation becomes
exact at $\s \simeq 0.06$. The ${\cal O}(\al)$ term that plays the
major role in changing the $\mu_b$-dependence of $A^c_9$ (see table 4)
and in removing this cancellation is proportional to the product of
$C^{c(1)}_1(\mu_0) = -15-6L\;$ from the matching conditions and $\;\ln
m_b/\mu_b\;$ from the one-loop matrix element of $P^c_1$. A future
calculation of the two-loop $b \to s l^+ l^-$ matrix elements of the
four-quark operators is desirable, because it should significantly
reduce the $\mu_b$-dependence of the prediction for
$R^{l^+l^-}_{quark}(\s)$.

When the results described by the solid lines in fig.~\ref{mubdep} are
integrated over $\s$, we obtain
\be \label{numR}
\int_{0.05}^{0.25} d\s \; R^{l^+l^-}_{quark}(\s) = (1.36 \pm 0.18)\times 10^{-5},
\ee
where only the error from $\mu_b$-dependence is taken into account.
Varying $U_{92}^{c(2)}$ from $-10$ to $10$ (as promised at the end of
the previous section) would increase the uncertainty by only 0.03.
Thus, calculating the three-loop anomalous dimensions in the future is
not expected to have an important impact on the numerical prediction.

In the end, we relate the integrand of $R^{l^+l^-}_{quark}(\s)$ to the
physically measurable quantity 
\bea 
BR[ B \to X_s l^+ l^-]_{\s \in [0.05, 0.25]} &=&
BR[ B \to X_c e \bar{\nu}] \; \int_{0.05}^{0.25} d\s \left[ R^{l^+l^-}_{quark}(\s)
+ \delta_{1/m_c^2} R(\s) + \delta_{1/m_b^2} R(\s) \right]
\nonumber\\[2mm]&&\hspace{-3cm}
= 0.104 [ (1.36 \pm 0.18) - 0.02 + 0.06] \times 10^{-5}
= (1.46 \pm 0.19) \times 10^{-6}, \label{numfinal}
\eea 
where, again, only the error from the $\mu_b$-dependence of
$R^{l^+l^-}_{quark}(\s)$ is included.  The non-perturbative HQET
corrections $\delta_{1/m_c^2} R(\s)$ and $\delta_{1/m_b^2} R(\s)$ have
been found with the help of eq.~(32) in ref.~\cite{BIR98} and eq.~(18) in
ref.~\cite{BI98}, respectively. The ${\cal O}(1/m_b^3)$ effects are
completely negligible for $\s < 0.25$ \cite{BaBu99}. The experimental
value of 0.104 for the semileptonic branching ratio is taken from
ref.~\cite{PData98}.

It is worth indicating that additional non-perturbative corrections due
to the motion of the $b$-quark inside the B-meson would occur if we
wanted to impose additional cuts on the emitted lepton energies
\cite{AH98}. Such corrections are absent only when the kinematical cut
is imposed on nothing but the invariant mass of the lepton pair.

Of course, translating the restriction $\s \in [0.05,0.25]$ to bounds
in GeV on the lepton invariant mass introduces an additional
uncertainty due to the numerical value of $m_{b,pole}$. Since the
$\s$-spectrum is almost flat in the considered domain, this additional
uncertainty (in per cent) will be close to $\f{5}{4}
\sigma_{m_{b,pole}}/m_{b,pole}$, i.e. rather small.

Finally let us note that restricting the studied domain of $\s$ to
$[0.05,0.25]$ makes the integrated $B \to X_s l^+l^-$ branching ratio
smaller, but at the same time more sensitive to the sign of
$\tilde{C}_7^{eff}(\mu_b)$, when compared to the so-called
``non-resonant BR'' considered for instance in ref.~\cite{CMW96}. If
we changed the sign of $\tilde{C}_7^{eff}(\mu_b)$, the last result in
eq.~(\ref{numfinal}) would change to $2.92 \times 10^{-6}$. Thus,
extensions of the SM that predict opposite sign of
$\tilde{C}_7^{eff}(\mu_b)$ (like the MSSM in certain
dark-matter-favoured regions of its parameter space) might be tested
with the help of the integrated BR itself, without considering
forward--backward or energy asymmetries.

At this point, we finish our phenomenological discussion, and proceed
to describing technical details of the two-loop matching computation
in the next section.

\section{Two-loop matching for photonic $\Delta B = - \Delta S= 1$\\
penguins in the Standard Model}
\label{details}

{\large \bf 5.1. Preliminaries}

For processes taking place at energy scales much lower than $M_W$, the
Standard Model can be replaced by an effective theory built out of
only light SM fields, i.e. the ones that are much lighter that the
W-boson. Our goal here is to find two-loop QCD contributions to the
Wilson coefficients of certain operators in the effective theory. The
operators we are interested in are the ones giving leading electroweak
contributions to the $\Delta B = - \Delta S= 1$ transitions
accompanied by either a real photon or a lepton pair emission. In the
latter case, we restrict ourselves to processes mediated by a virtual
photon, i.e.  we do not consider in this section the SM diagrams where
the W or Z boson couple directly to the lepton line.

The simplest way to find the Wilson coefficients is to require
equality of the off-shell 1PI amputated Green functions calculated in
the full SM and in the effective theory. Up to one loop,
we need to consider the $b \to s \gamma$, $b \to s\;gluon$ and $b \to
s c \bar{c}$ functions. At two loops, only the $b \to s \gamma$
function is necessary. In the cases of $b \to s \gamma$ and $b \to
s\;gluon$, we work at the leading order in $\alpha_{em}$ and up to
${\cal O}[($external momenta$)^2/M_W^2]$.  In the $b \to s c \bar{c}$
case, external momenta can be neglected.

We set all the light particle masses to zero in the whole calculation.
An exception is the $b$-quark mass, which is being included up to
linear order. This means that we maintain $m_b$ only in Yukawa
couplings and in the $b$-quark propagator numerators. The terms of
order $m_b^2$ are neglected. One can justify this procedure by
formally treating the $b$-quark mass term as an interaction with an
external scalar field.

        In addition, all the Feynman integrands are expanded in
external momenta before performing loop integration. Such an
expansion, as well as setting all the light masses to zero, creates
spurious infrared divergences that we regularize dimensionally. As we
shall see, all these divergences cancel out in the matching conditions
relating the full and the effective theory Green functions.

The Feynman integrands for the one- and two-loop Feynman diagrams are
generated with the help of the program {\it FeynArts} \cite{KBD90}.
After Taylor expansion in external momenta and factorizing them out,
the integrals remain dependent only on loop momenta and two heavy
masses: $M_W$ and $m_t$. Subsequent application of the partial
fraction decomposition
\be
\f{1}{(q^2-m_1^2)(q^2-m_2^2)} = 
\f{1}{m_1^2-m_2^2} \left[ \f{1}{q^2-m_1^2} - \f{1}{q^2-m_2^2} \right]  
\ee
allows a reduction of all the integrals to those in which a single
mass parameter occurs in the propagator denominators together with a
given loop momentum. Finally, after reduction of tensor integrals to
scalar ones, the non-vanishing integrals obtained at one and two loops
are respectively as follows:
\bea
C^{(1)}_n &=& \f{(m^2)^{n-2+\e}}{\pi^{2-\e} \; \Gamma(1+\e)}
              \int \f{d^{4-2\e}\;q}{(q^2 - m^2)^n},
\label{int1}\\[2mm]
C^{(2)}_{n_1 n_2 n_3} &=& 
\f{(m_1^2)^{n_1+n_2+n_3-4+2\e}}{\pi^{4-2\e}\; \Gamma(1+\e)^2}
\int \f{d^{4-2\e}q_1 \; d^{4-2\e}q_2}{(q_1^2 - m_1^2)^{n_1}
                  (q_2^2 - m_2^2)^{n_2}[(q_1 - q_2)^2]^{n_3}},
\label{int2}
\eea
with arbitrary integer powers $n$, $n_1$, $n_2$ and $n_3$, and with
$m,\;m_1 \neq 0$. The chosen normalization makes the results free of
trivial common factors.

In eq.~(\ref{int2}) we have already made use of the fact that our
two-loop scalar integrals always have at least one massless term in
their denominators. This turns out to be true in all the Feynman
diagrams we have to consider, provided all the light particle masses
are set to zero. Therefore, all our two-loop integrals are relatively
simple.

        The result for the one-loop scalar integral is
\be
C^{(1)}_n = i \f{(-1)^n}{(n-1)!} (1+\e)_{n-3},
\ee
which vanishes for $n \leq 0$. Here, $(a)_k$ denotes the Pochhammer
symbol equal to
\be
(a)_k = \f{\Gamma(a+k)}{\Gamma(a)}= \left\{ \begin{array}{cc}
a(a+1)(a+2)...(a+k-1),    & k \geq 1,\\
1,                        & k = 0,\\
1/[(a-1)(a-2)...(a-|k|)], & k \leq -1,
\end{array} \right.
\ee
for integer $k$ and complex $a$.

The two-loop integrals can easily be found with the help of Feynman
parametrization in the cases when $m_1 = m_2$~ or ~$m_2 = 0$
\bea \label{page55}
C^{(2)}_{n_1 n_2 n_3} &\bbuildrel{=\!=\!=\!=}_{\scs m_1 = m_2}^{}&
(-1)^{n_1+n_2+n_3+1} \f{(2-\e)_{-n_3} (1+\e)_{n_1+n_3-3}(1+\e)_{n_2+n_3-3}
}{(n_1-1)! (n_2-1)! (n_1+n_2+n_3-4+2\e)_{n_3}},
\\[1mm] \label{page68}
C^{(2)}_{n_1 n_2 n_3} &\bbuildrel{=\!=\!=\!=}_{\scs m_2 = 0}^{}&
(-1)^{n_1+n_2+n_3+1} 
\f{ (1+2\e)_{n_1+n_2+n_3-5} (1+\e)_{n_2+n_3-3} (1-\e)_{1-n_2} (1-\e)_{1-n_3}
}{ (n_1-1)! (n_2-1)! (n_3-1)! (1-\e)(1-\f{1}{3}\pi^2\e^2 + {\cal O}(\e^3))}.
\hspace{1cm}
\eea
It remains to discuss the case when $m_1 \neq m_2$ and none of the two
masses vanishes. The starting point is the integral $C^{(2)}_{111}$,
which equals:
\be
C^{(2)}_{111} = 
\f{1}{2 (1-\e) (1-2\e)} \left[ -\f{1+x}{\e^2} \;+\; \f{2}{\e} x \ln x \;+\;
(1-2x) \ln^2 x \;+\; 2 (1-x) Li_2\left(1-\f{1}{x}\right) \;+\; {\cal O}(\e) \right],
\ee
where $x = m_2^2/m_1^2$ \cite{DT93}. All the integrals with three
positive indices can be derived from the above result with the help of
the following recurrence relations \cite{DT93}:
\be
\begin{array}{ccrlrl} 
C^{(2)}_{(n_1+1) n_2 n_3} &=& \f{1}{n_1 (1-x)} 
& \multicolumn{3}{l}{  
\left\{ [ 4-2\e-n_1-n_2-n_3 + x (n_1 -n_3)] C^{(2)}_{n_1 n_2 n_3}
\right.} \vspace{0.2cm} \\ &&&  
+ &       x n_2 & \left. \left[ C^{(2)}_{(n_1-1)(n_2+1)n_3} 
                               -C^{(2)}_{n_1(n_2+1)(n_3-1)} \right] \right\},
\\[2mm] 
C^{(2)}_{n_1 (n_2+1) n_3} &=& -\f{1}{n_2 x (1-x)} 
& \multicolumn{3}{l}{
\left\{ [ x (4-2\e-n_1-n_2-n_3) + n_2 -n_3] C^{(2)}_{n_1 n_2 n_3}
\right.} \vspace{0.2cm} \\ &&&   
  + &       n_1 & \left. \left[ C^{(2)}_{(n_1+1)(n_2-1)n_3} 
                               -C^{(2)}_{(n_1+1)n_2(n_3-1)} \right] \right\},
\\[2mm] 
C^{(2)}_{n_1 n_2 (n_3+1)} &=& \f{1}{n_3 (1-x)^2} 
& \multicolumn{3}{l}{
\left\{ [(1+x)(-4+2\e) + 2 n_2 + (1+3x) n_3] C^{(2)}_{n_1 n_2 n_3}
\right.} \vspace{0.2cm} \\ &&&  
  + &  2 x n_2  &        \left[ C^{(2)}_{n_1(n_2+1)(n_3-1)} 
                               -C^{(2)}_{(n_1-1)(n_2+1)n_3} \right]
\vspace{0.2cm} \\ &&& 
  + & (1-x) n_3 & \left. \left[ C^{(2)}_{n_1(n_2-1)(n_3+1)}   
                               -C^{(2)}_{(n_1-1)n_2(n_3+1)} \right] \right\}.
\end{array}
\ee

        All the two-loop integrals defined in eq.~(\ref{int2}) vanish
when either $n_1$ or $n_2$ is non-positive. When these two indices are
positive but $n_3$ is non-positive, they reduce to products of one-loop
tensor integrals. It is sensible to make this reduction only in the
case when the two masses are different and non-vanishing. Then we obtain
\bea 
C^{(2)}_{n_1 n_2 n_3} &\bbuildrel{=\!=\!=\!=}_{\scs n_3 \leq 0}^{}&
\sum_{k=0}^{[-\f{n_3}{2}]} \sum_{j=0}^{-n_3-2k} 
\newton{-n_3}{2k} \newton{-n_3-2k}{j} \f{ x^{2-n_2+k+j-\e} 
(-1)^{n_1+n_2+n_3+1} (2k)!}{(n_1-1)! (n_2-1)! k! (2-\e)_k} 
\times \nonumber \\ \nonumber \\ \label{page54} && \times 
(2-\e)_{j+k} (2-\e)_{-n_3-k-j} (1+\e)_{n_2-k-j-3} (1+\e)_{n_1+n_3+k+j-3}.
\eea
Otherwise, one can use eqs. (\ref{page55}) and (\ref{page68}), which
apply for non-positive $n_3$, too. Equation (\ref{page68}) gives zero
in such a case, but eq. (\ref{page55}) does not.

\ \\ \
{\large \bf 5.2. The Standard Model side}

\begin{figure}[h]
\vspace{6mm}
\hspace*{11mm}  $\gamma$ \hspace{36.5mm} $\gamma$ 
\hspace{41.5mm} $\gamma$ \hspace{36.5mm} $\gamma$ \\[7mm] 
\hspace*{0mm} $u,c,t$ \hspace{9mm}     $u,c,t$ \hspace{12mm} 
            $W^{\pm}$ \hspace{8mm}   $W^{\pm}$ \hspace{15mm}
              $u,c,t$ \hspace{9mm}     $u,c,t$ \hspace{14mm} 
          $\pi^{\pm}$ \hspace{8mm} $\pi^{\pm}$ \\[-18mm] 
\includegraphics[width=75mm,angle=0]{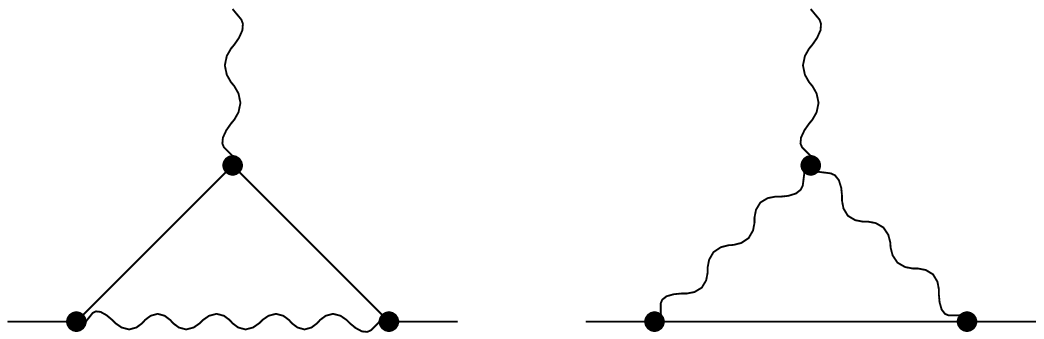}
\hspace{1cm}
\includegraphics[width=75mm,angle=0]{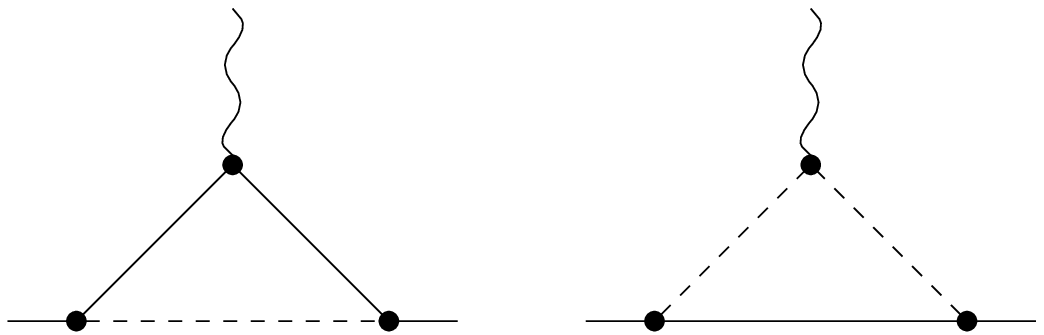}\\[-2mm]
$b$ \hspace{1cm}  $W^{\pm}$  \hspace{7mm} $s$ \hspace{6mm} 
$b$ \hspace{7mm}   $u,c,t$   \hspace{7mm} $s$ \hspace{11mm} 
$b$ \hspace{1cm} $\pi^{\pm}$ \hspace{8mm} $s$ \hspace{6.5mm} 
$b$ \hspace{7mm}   $u,c,t$   \hspace{7mm} $s$ \\[-1cm]
\begin{center}
\caption{One-loop 1PI diagrams for $b \to s \gamma$ in the SM.
         The charged would-be Goldstone boson is denoted by $\pi^{\pm}$.
         There is no $W^{\pm}\pi^{\mp}\gamma$ coupling in the background-field gauge.}
\label{1loop.bsgamma}
\end{center}
\vspace{-1cm}
\end{figure}

        Let us start with calculating the $b \to s \gamma$ function up
to two loops.  There is no tree-level contribution to this function in
the Standard Model.  The four 1PI diagrams arising at one loop are
presented in fig.~\ref{1loop.bsgamma}.

We calculate the corresponding unrenormalized amputated Green function
off shell, in the background-field version of the 't~Hooft-Feynman
gauge. The Feynman integrands are expanded up to the second order in
external momenta and $m_b$ (neglecting $m_b^2$ though). As in
section~\ref{matching}, we refrain from using unitarity of the CKM
matrix here. The result can be written in the following form:
\be \label{green1.bsgamma}
i \f{4 G_F}{\sqrt{2}} \f{e P_R}{(4 \pi)^2} N^{(1)}_{\e}
\left\{ \left( V_{us}^* V_{ub} + V_{cs}^* V_{cb} \right) \sum_{j=1}^{13} h^{(1)}_j S_j
\;\; + \;\;  V_{ts}^* V_{tb} \sum_{j=1}^{13} f^{(1)}_j(x) S_j \right\}
\;   + \;   {\cal O}(\e^2),
\ee
where 
$P_R = \f{1}{2}(1+\gamma_5)$,~ 
$N^{(1)}_{\e} = 1 - \e \kappa   + \e^2 ( \f{1}{12} \pi^2 + \f{1}{2} \kappa^2)$,~ 
$\kappa = \gamma_{\scs E} - \ln(4 \pi) + \ln(M_W^2/\mu_0^2)$~ 
and $S_k$ stand for Dirac structures that depend
on the incoming $b$-quark momentum $p$ and on the outgoing photon momentum $k$
\bea
S_j &=& \left( \gamma_{\mu} \slash p \slash k, \; 
               \gamma_{\mu} \; (p \cdot k), \;
               \gamma_{\mu} p^2, \;
               \gamma_{\mu} k^2, \;
               \slash p k_{\mu}, \;
               \slash p p_{\mu}, \;
               \slash k p_{\mu}, \;
               \slash k k_{\mu}, \;
\right. \nonumber\\ && \hspace{4cm} \left. 
               m_b \slash k \gamma_{\mu}, \;
               m_b \gamma_{\mu} \slash k, \;
               m_b \slash p \gamma_{\mu}, \;
               m_b \gamma_{\mu} \slash p, \;
               M_W^2 \gamma_{\mu} \right)_j.
\eea
As we shall see later, explicit results are needed only for the
coefficients at the structures $S_2$, $S_8$ and $S_{10}$. We find
\be
\begin{array}{rcl}
h^{(1)}_2 &=& \f{23}{9} + \f{145}{54} \e,
\hspace{3cm}  
h^{(1)}_8 = -\f{4}{9\e} + \f{7}{54} + \f{59}{324} \e,
\hspace{3cm}  
h^{(1)}_{10} = 0,\\[3mm]
f^{(1)}_2(x) &=& 
  \f{15x^3-16x^2+4x}{3(x-1)^4} \ln x 
+ \f{-8x^3-105x^2+141x-46}{18(x-1)^3}\\[2mm]
&& +\e \; \left\{ 
  \f{-15x^3+16x^2-4x}{6(x-1)^4} \ln^2 x
+ \f{8x^4+115x^3-150x^2+48x}{18(x-1)^4} \ln x 
+ \f{-76x^3-645x^2+885x-290}{108(x-1)^3} \right\},\\[4mm]
f^{(1)}_8(x) &=&    
  \f{-3x^4-15x^3-6x^2+20x-8 }{18(x-1)^4} \ln x
+ \f{71x^3+78x^2-111x+34}{108(x-1)^3}
  +\e \; \left\{ 
  \f{3x^4+15x^3+6x^2-20x+8}{36(x-1)^4} \ln^2 x
\right. \\[2mm] && \left.
+ \f{-71x^4-79x^3+162x^2-144x+48}{108(x-1)^4} \ln x
+ \f{529x^3-102x^2+195x-118}{648(x-1)^3} \right\},\\[4mm]
f^{(1)}_{10}(x) &=&    
  \f{-3x^2+2x}{6(x-1)^3}  \ln x
+ \f{5x^2-3x}{12(1 - x)^2}
 +\e \; \left\{ 
  \f{3x^2-2x}{12(x-1)^3}  \ln^2 x
+ \f{-5x^3+2x^2}{12(x-1)^3} \ln x
+ \f{11x^2-5x}{24(x-1)^2} \right\}, 
\end{array} \label{1loop.bsgamma.sum}
\ee
where $x = m_t^2/M_W^2$. 

\begin{figure}[h]
\includegraphics[width=75mm,angle=0]{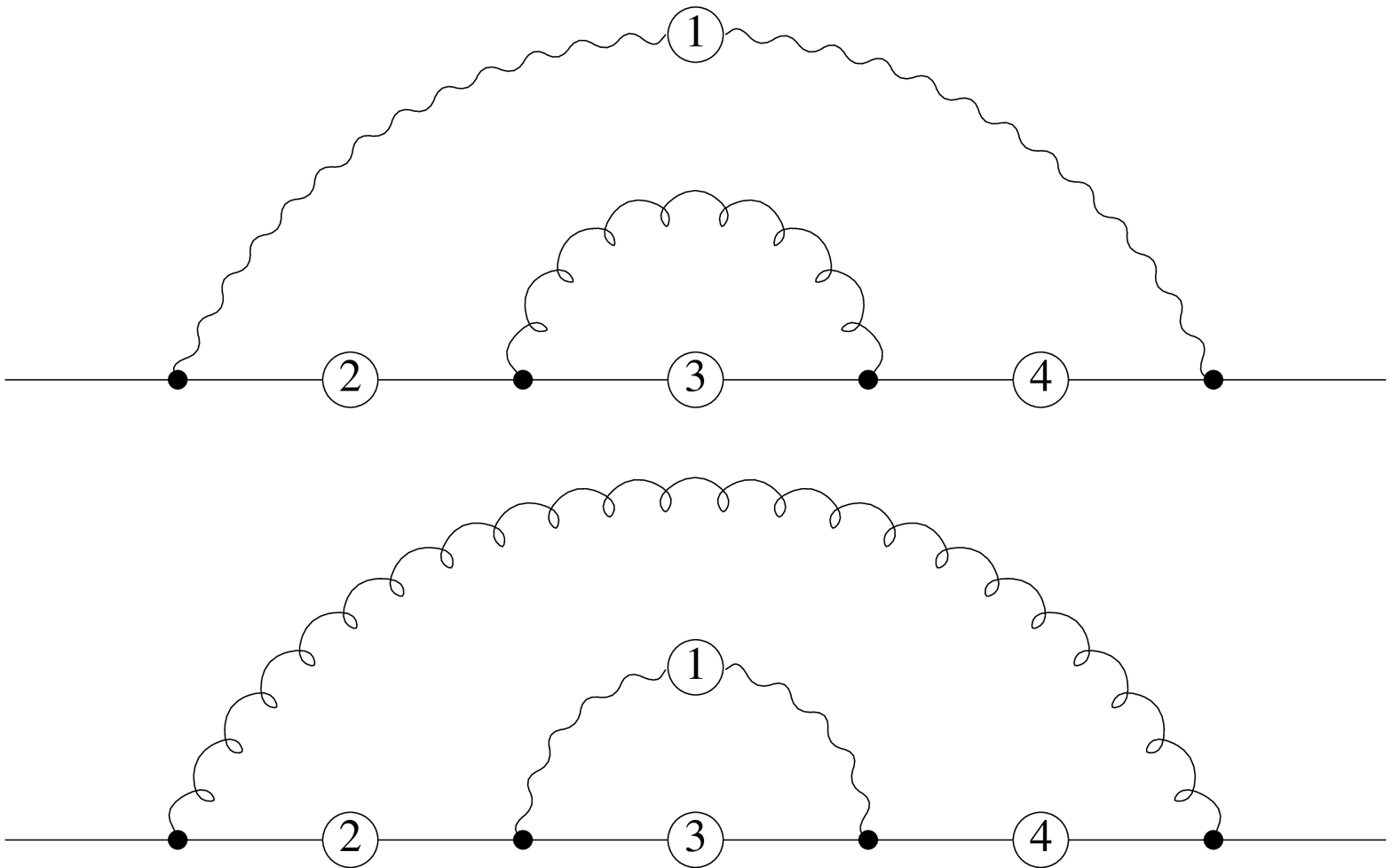}
\hspace{1cm}
\includegraphics[width=75mm,angle=0]{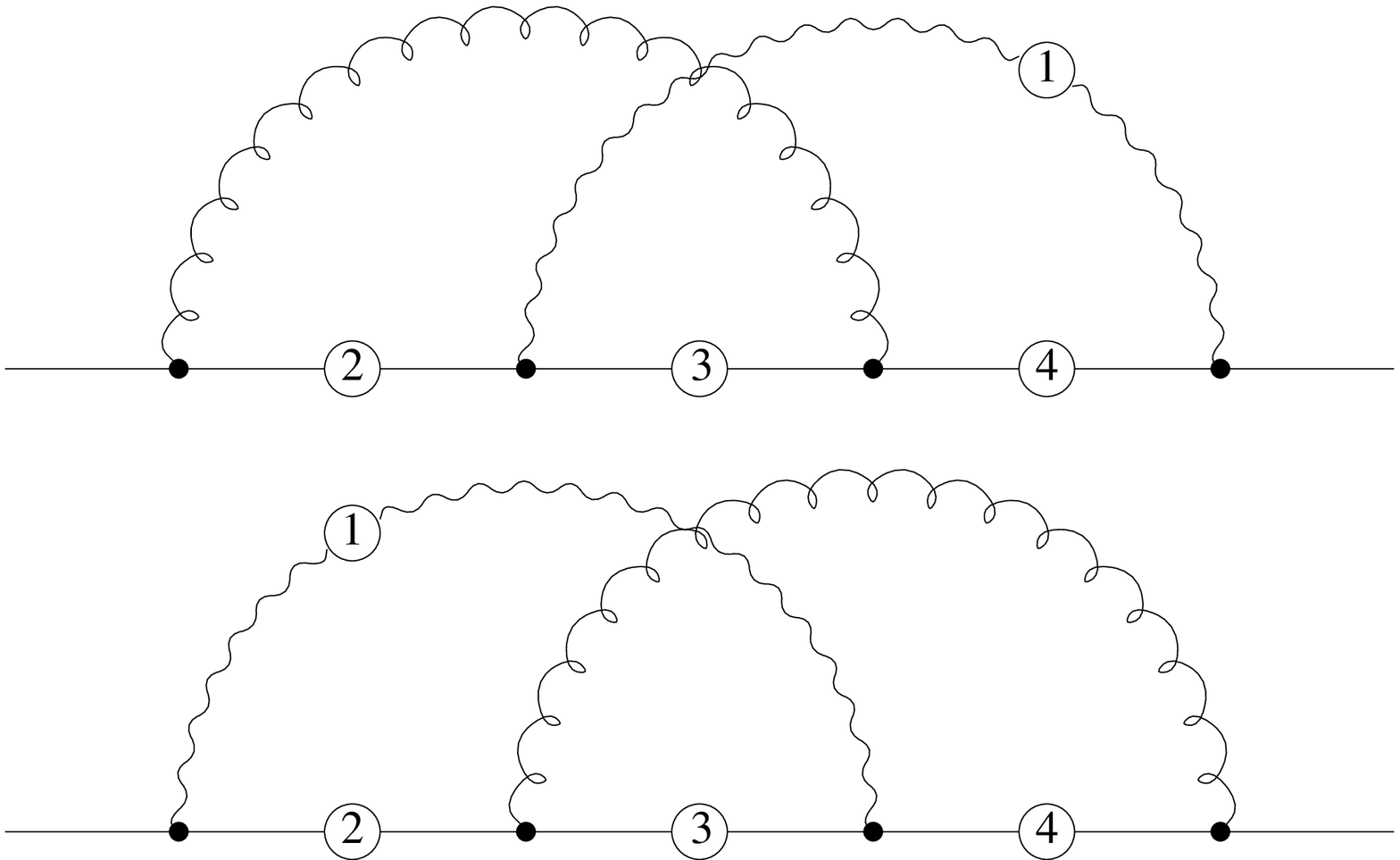}
\begin{center}
\caption{Two-loop 1PI diagrams for $b \to s \gamma$ in the SM. The wavy lines
         denote either the $W$-boson or the charged would-be Goldstone boson. 
         The external photon can couple at any of the places marked by small circles.}
\label{2loop.bsgamma}
\end{center}
\vspace{-1cm}
\end{figure}

Let us now proceed to an evaluation of the first QCD correction to the
considered Green function. The corresponding two-loop diagrams are
shown in fig.~\ref{2loop.bsgamma}.

In analogy to eq.~(\ref{green1.bsgamma}), we write the unrenormalized
two-loop result as 
\be \label{green2.bsgamma}
i \f{4 G_F}{\sqrt{2}} \f{e g^2 P_R}{(4 \pi)^4} N^{(2)}_{\e}
\left\{ \left( V_{us}^* V_{ub} + V_{cs}^* V_{cb} \right) \sum_{j=1}^{13} h^{(2)}_j S_j
\;\; + \;\;  V_{ts}^* V_{tb} \sum_{j=1}^{13} f^{(2)}_j(x) S_j \right\}
\;   + \;   {\cal O}(\e),
\ee
where $g$ is the QCD gauge coupling and 
$N^{(2)}_{\e} = 1 - 2 \e \kappa + \e^2 ( \f{1}{6} \pi^2 + 2 \kappa^2)$. 
The two-loop analogues of the coefficients given in
eq.~(\ref{1loop.bsgamma.sum}) are found to have the following form:
\bea
\begin{array}{rcl} 
h^{(2)}_2 &=& -\f{272}{81\e}-\f{3740}{243}, 
\hspace{1.5cm}
h^{(2)}_8 = -\f{128}{81\e^2} - \f{1088}{243\e} -\f{314}{729} - \f{128\pi^2}{243},
\hspace{1.5cm}
h^{(2)}_{10} = \f{20}{9\e} +\f{92}{27},
\\[4mm]
f^{(2)}_2(x) &=& \f{1}{\e} \left\{
  \f{8x(-45x^3-34x^2+53x-10)}{9(x-1)^5} \ln x
+ \f{4(x^4+641x^3-501x^2+83x-8)}{27(x-1)^4} \right\} 
\\[2mm]&& 
+ \f{8x(7x^3-69x^2+61x-14)}{9(x-1)^4} Li_2\left( 1-\f{1}{x} \right) 
+ \f{4x(45x^3+34x^2-53x+10)}{3(x-1)^5} \ln^2 x 
\\[2mm]&& 
+ \f{4(-6x^5-4497x^4+2622x^3+811x^2-638x+88)}{81(x-1)^5} \ln x
+ \f{2(-719x^4+35822x^3-35073x^2+11492x-1802)}{243(x-1)^4},
\end{array} \nonumber \eea \be \begin{array}{rcl} 
f^{(2)}_8(x) &=& \f{1}{\e} \left\{
\f{4(243x^4+486x^3-419x^2+130x-8)}{81(x-1)^5} \ln x
+ \f{2(-185x^4-3313x^3+369x^2+905x-368)}{243(x-1)^4} \right\} 
\\[2mm]&&
+ \f{4(32x^4+283x^3-135x^2-70x+64)}{81(x-1)^4} Li_2\left( 1-\f{1}{x} \right)
+ \f{2(-243x^4-486x^3+419x^2-130x+8)}{27(x-1)^5} \ln^2 x
\\[2mm]&&
+ \f{2(370x^5+7933x^4-1370x^3-683x^2+238x-8)}{243(x-1)^5} \ln x
+ \f{2(-3301x^4-20714x^3+4182x^2+202x+191)}{729(x-1)^4},\\[4mm]
f^{(2)}_{10}(x) &=& \f{1}{\e} \left\{
  \f{2x(36x^2+x-10)}{9(x-1)^4} \ln x 
+ \f{11x^3-169x^2+132x-28}{9(x-1)^3} \right\} 
+ \f{2x(-15x^3+8x^2-21x+10)}{9(x-1)^4} Li_2\left( 1-\f{1}{x} \right)
\\[2mm]&&
+ \f{x(-36x^2-x+10)}{3(x-1)^4} \ln^2 x
+ \f{-22x^4+396x^3-377x^2+142x-16}{9(x-1)^4} \ln x
+ \f{31x^3-1071x^2+630x-112}{54(x-1)^3}.
\end{array}
\ee

        The last two elements we need to know on the SM
side are the $b \to s\;gluon$ and $b \to s c \bar{c}$ functions up to
one loop. They are used to recover one-loop contributions to certain
Wilson coefficients which take part in the two-loop $b \to s \gamma$
matching condition.

\begin{figure}[h]
\vspace{15mm}
\begin{center}
$u,c,t$ \hspace{9mm}     $u,c,t$ \hspace{8.5mm} 
$u,c,t$ \hspace{9mm}     $u,c,t$\\[-17.5mm] 
\includegraphics[width=75mm,angle=0]{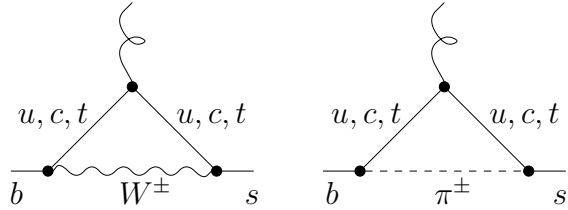}\\[-2mm]
$b$ \hspace{1cm}  $W^{\pm}$  \hspace{7mm} $s$ \hspace{6mm} 
$b$ \hspace{1cm} $\pi^{\pm}$ \hspace{8mm} $s$ \\
\caption{One-loop 1PI diagrams for $b \to s\;gluon$ in the SM.}
\label{1loop.bsgluon}
\end{center}
\vspace{-5mm}
\end{figure}
Similarly to the $b \to s \gamma$ case, there is no tree-level
contribution to the $b \to s\;gluon$ Green function in the SM. The
one-loop contribution is given by the two diagrams presented in
fig.~\ref{1loop.bsgluon}. In analogy to eq.~(\ref{green1.bsgamma}),
the result can be written as
\be \label{green1.bsgluon}
i \f{4 G_F}{\sqrt{2}} \f{g P_R T^a}{(4 \pi)^2} N^{(1)}_{\e}
\left\{ \left( V_{us}^* V_{ub} + V_{cs}^* V_{cb} \right) \sum_{j=1}^{13} u^{(1)}_j S_j
\;\; + \;\;  V_{ts}^* V_{tb} \sum_{j=1}^{13} v^{(1)}_j(x) S_j \right\}
\;   + \;   {\cal O}(\e^2),
\ee
where $T^a$ denotes the SU(3) generator corresponding to the outgoing
gluon. The coefficients at the structures $S_2$, $S_8$ and $S_{10}$
read
\be
\begin{array}{rcl} 
u^{(1)}_2 &=& \f{4}{3} + \f{22}{9} \e,
\hspace{2cm}
u^{(1)}_8 = -\f{2}{3\e} + \f{1}{9} + \f{11}{54} \e,
\hspace{2cm}
u^{(1)}_{10} = 0,
\end{array}
\ee
\be
\begin{array}{rcl} 
v^{(1)}_2(x) &=& \f{-5x^2+2x}{(x - 1)^4} \ln x
+ \f{-x^3+15x^2+12x-8}{6(x-1)^3} 
\\[2mm]&&
+\e \; \left\{ 
  \f{5x^2-2x}{2(x-1)^4} \ln^2 x
+ \f{x^4-16x^3-30x^2+24x}{6(x - 1)^4} \ln x
+ \f{-5x^3+159x^2+60x-88}{36(x - 1)^3} \right\},
\\[4mm]
v^{(1)}_8(x) &=& \f{3x^2+5x-2}{3(x-1)^4} \ln x
+ \f{5x^3-12x^2-39x+10}{18(x-1)^3}  
\\[2mm]&&
+\e \; \left\{ 
  \f{-3x^2-5x+2}{6(x-1)^4} \ln^2x
+ \f{-5x^4+17x^3+54x^2-36x+12}{18(x-1)^4} \ln x 
+ \f{19x^3-192x^2-57x-22}{108(x-1)^3} \right\},
\\[4mm]
v^{(1)}_{10}(x) &=& \f{x}{2(x-1)^3} \ln x
+ \f{x^2-3x}{4(x-1)^2} 
+\e \; \left\{ 
  \f{-x}{4(x-1)^3} \ln^2 x
+ \f{-x^3+4x^2}{4(x-1)^3} \ln x
+ \f{x^2-7x}{8(x-1)^2} \right\}.
\end{array}
\ee

\begin{figure}[h]
\begin{center}
\includegraphics[width=3cm,angle=0]{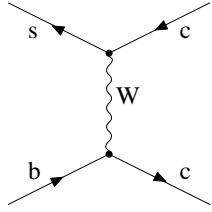}
\caption{Tree-level $b \to s c \bar{c}$ diagram on the SM side.}  
\label{tree.bscc}
\end{center}
\end{figure}
        Contrary to the functions considered so far, the $b \to s c
\bar{c}$ function does acquire a tree-level contribution in the
SM. It is given by the diagram shown in fig.~\ref{tree.bscc}. 
For vanishing external momenta, it gives\footnote{
The tensor product symbol $\Gamma \otimes \Gamma'$ is used here to
denote the tree-level $(\bar{s} \Gamma c)(\bar{c} \Gamma' b)$
amputated Green function.}
\be \label{green0.bscc}
- i \f{4 G_F}{\sqrt{2}} V_{cs}^* V_{cb} (\gamma_{\mu} P_L) \otimes (\gamma^{\mu} P_L ).
\ee
\begin{figure}[h]
\begin{center}
\includegraphics[width=12cm,angle=0]{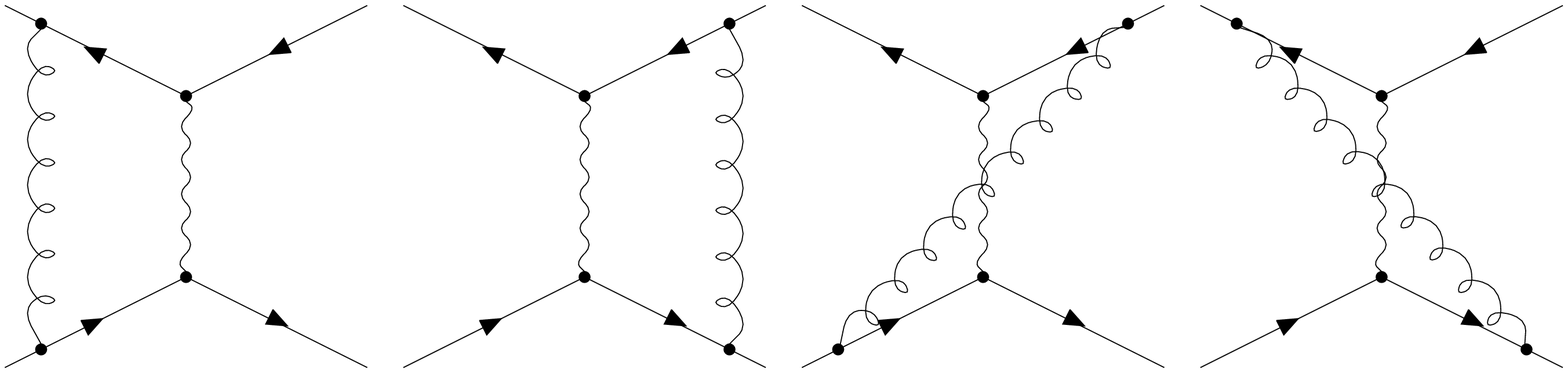}
\caption{One-loop $b \to s c \bar{c}$ diagrams on the SM
side, which do not vanish in dimensional regularization when all the
light particle masses are set to zero. \label{1loop.bscc}}
\end{center}
\end{figure}

The non-vanishing one-loop diagrams for the $b \to s c \bar{c}$
functions are shown in fig.~\ref{1loop.bscc}. When the external
momenta are set to zero, we find the following result for the
corresponding amputated Green function:
\bea
i \f{4 G_F}{\sqrt{2}} \f{g^2}{(4 \pi)^2} V_{cs}^* V_{cb} N^{(1)}_{\e}
\left\{ \left( -\f{6}{\e} -15 -\f{39}{2} \e \right) 
   (\gamma_{\mu} P_L T^a) \otimes (\gamma^{\mu} P_L T^a) 
\;\; + \;\; \left( -\f{1}{\e} -\f{3}{2} +{\cal O}(\e) \right) 
\times \right. \nonumber \\[-2mm] \nonumber 
\eea
\be
\left. \times \left[  (\gamma_{\mu} \gamma_{\nu} \gamma_{\rho} P_L T^a) 
\otimes (\gamma^{\mu} \gamma^{\nu} \gamma^{\rho} P_L T^a)
-16 (\gamma_{\mu} P_L T^a) \otimes (\gamma^{\mu} P_L T^a) \right]
\right\} \; + \;   {\cal O}(\e^2).
\label{green1.bscc}
\ee

        The Dirac structure in the last line of the above equation
vanishes in four dimensions. However, there is no way to express it as
$\e \times$(simpler structure). The coefficient at this structure will
give us the Wilson coefficient of an evanescent operator in the
effective theory \cite{evan}. The necessity of recovering this
coefficient (as well as keeping ${\cal O}(\e)$ parts of other one-loop
coefficients) is a price we have to pay for regularizing infrared
divergences dimensionally.

        The above result is the last one we need to know on the
SM side. In the next subsection, we shall study the same
Green functions in the effective theory framework.

\ \\
{\large \bf 5.3. The effective theory side}

The lagrangian of the effective theory has been given in
eq.~(\ref{Leff}). At present, we need to include in addition several
non-physical operators. We write
\bea 
{\cal L}_{eff} &=& {\cal L}_{\scs QCD \times QED}(u,d,s,c,b,e,\mu,\tau) 
+ \f{4 G_F}{\sqrt{2}} \left\{
\sum_{Q=u,c} V^*_{Qs} V_{Qb} (C^c_1 P^Q_1 + C^c_2 P^Q_2 + C^c_{11} P^Q_{11}) 
\right. \nonumber \\ && \hspace{3cm} \left.
+\sum_i  
[(V^*_{us} V_{ub} + V^*_{cs} V_{cb}) C^c_i \; + \; V^*_{ts} V_{tb} C^t_i] P_i
\right\}. \label{xLeff}
\eea

The operators $P^Q_i$ and $P_i$ entering the effective lagrangian can
be divided into three classes: physical, evanescent (i.e.
algebraically vanishing in four dimensions) and EOM-vanishing (i.e.
vanishing by the QCD$\times$QED equations of motion, up to a total
derivative).

The physical operators have already been given in
eq.~(\ref{physical}). However, for the purpose of the present section,
it is convenient to redefine $P_9$ so that it contains a sum over all
the light charged fermions $f$ weighted by their electric charges $Q_f$
\be \label{P9new}
P_9 =  -\f{e^2}{g^2} (\bar{s}_L \gamma_{\mu} b_L) \sum_f Q_f (\bar{f}\gamma^{\mu} f).
\ee
Such a redefinition of $P_9$ does not alter its Wilson coefficient at
leading order in electroweak interactions.

As far as the evanescent operators are concerned, only $P^Q_{11}$ from
the appendix will be needed in the present section.

The gauge-invariant EOM-vanishing operators can be chosen as 
\bea \label{EOM-vanishing}
P_{31} &=& \f{1}{g} (\bar{s}_L \gamma^{\mu} T^a b_L) 
                    D^{\nu} G^a_{\mu \nu} \; + \; P_4,
\nonumber\\
P_{32} &=& \f{1}{g^2} m_b \bar{s}_L \slash D \slash D b_R,
\nonumber\\
P_{33} &=& \f{i}{g^2} \bar{s}_L \slash D \slash D \slash D b_L,
\nonumber\\
P_{34} &=& \f{i}{g} \left[ \bar{s}_L \stackrel{\leftarrow}{\slash D} \sigma^{\mu \nu} T^a 
b_L G^a_{\mu \nu} -G^a_{\mu \nu} \bar{s}_L T^a \sigma^{\mu \nu} \slash D b_L \right] + P_8,
\nonumber\\
P_{35} &=& \f{ie}{g^2} \left[ \bar{s}_L \stackrel{\leftarrow}{\slash D} \sigma^{\mu \nu} 
b_L F_{\mu \nu} - F_{\mu \nu} \bar{s}_L \sigma^{\mu \nu} \slash D b_L \right] + P_7,
\nonumber\\
P_{36} &=& \f{e}{g^2} (\bar{s}_L \gamma^{\mu} b_L) 
                    \partial^{\nu} F_{\mu \nu} \; - \; P_9.
\eea
Our sign convention in the covariant derivative acting on a quark
field $\psi$ is
\be
D_{\mu} \psi = \left( \partial_{\mu} + i g G_{\mu}^a T^a + i e Q_{\psi}
A_{\mu} \right) \psi.
\ee

The EOM-vanishing operators in eq.~(\ref{EOM-vanishing}) can be
assumed to contain the background gluon field only, because nothing
but their tree-level matrix elements will be needed for the off-shell
matching in the next subsection. However, a systematic off-shell
renormalization of the effective theory requires introducing
EOM-vanishing operators that contain the quantum gluon field as well.
The explicit form of such operators is irrelevant here.  Nevertheless,
one should not forget that all of them enter into the sums over
operators, such as the one in the last term of eq.~(\ref{xLeff}).

It is not completely trivial to convince oneself that
eq.~(\ref{EOM-vanishing}) indeed contains all the gauge-invariant
EOM-vanishing operators that we may encounter. One way to do this is
to first write all the $\Delta B = -\Delta S = 1$ operators of
dimension 5 and 6 containing the left-handed $s$-quark field
only.\footnote{
  Here, the dimension of an operator is understood as the sum of
  dimensions of the fields and derivatives it contains. Explicit mass
  factors in the normalization are not counted.}
The derivatives acting on the $s$-quark field can be removed by parts.
One can start from writing down the 6 possible operators that contain
the chromomagnetic and electromagnetic field strength tensors or their
duals
\be 
\begin{array}{rrr}
(\bar{s}_L T^a \sigma^{\mu \nu}     b_R)                      G_{\mu \nu}^a, &
\hspace{2cm}
(\bar{s}_L T^a \gamma^{\mu}         b_L)              D^{\nu} G^a_{\mu \nu}, &
\hspace{2cm}
(\bar{s}_L T^a \gamma^{\mu} D^{\nu} b_L)               \tilde{G}^a_{\mu \nu}, \\[2mm]
(\bar{s}_L ~~~ \sigma^{\mu \nu}     b_R) \;                   F_{\mu \nu},   &
(\bar{s}_L ~~~ \gamma^{\mu}         b_L) \; \partial^{\nu} \; F_{\mu \nu},   &
(\bar{s}_L ~~~ \gamma^{\mu} D^{\nu} b_L) \;            \tilde{F}_{\mu \nu}.
\end{array}
\ee
Nothing new is obtained from the first two pairs of operators above,
when the field strength tensors are replaced by their duals, because
of the Bianchi identity and $\sigma_{\alpha\beta} \gamma_5 \sim
\varepsilon_{\alpha\beta\gamma\delta} \sigma^{\gamma\delta}$. On the
other hand, replacing the dual tensors by ordinary ones in the last
pair of operators would break CP combined with $b \leftrightarrow s$
interchange even for $m_b=0$ and real CKM angles.

The remaining operators (apart from the four-fermion ones) must
contain covariant derivatives. Since commutators of the covariant
derivatives give field strength tensors, only one additional operator
with three covariant derivatives (e.g. $\bar{s}_L \slash D D^2 b_L$)
and one operator with two covariant derivatives (e.g. $\bar{s}_L D^2
b_L$) remains. At this point, one has at hand a complete set of 8
gauge-invariant operators (apart from the four-fermion ones). The
``magnetic moment'' operators $P_7$, $P_8$ and the EOM-vanishing
operators $P_{31}$, ..., $P_{36}$ are just certain linear combinations
of them, $P_4$ and $P_9$ (up to total derivatives).

Since both the $u$- and $c$-quarks are treated as massless in the
present calculation, the lagrangian is symmetric under $u
\leftrightarrow c$ exchange. This symmetry has already been taken into
account in eq.~(\ref{xLeff}): the same Wilson coefficients $C^c_i$
occur both in the $u$-quark and the $c$-quark sectors.

        The lagrangian (\ref{xLeff}) is written in terms of bare fields
and parameters. In order to express it in terms of the QCD-renormalized
quantities, we replace 
\be
g     \to Z_g g, \hspace{18mm}
m_b   \to Z_m m_b, \hspace{18mm}
\psi  \to Z_{\psi}^{1/2} \psi, \hspace{18mm}
C^Q_i \to \sum_j C^Q_j Z_{ji}, 
\ee
for the QCD gauge coupling, $b$-quark mass, quark fields and the Wilson
coefficients, respectively. As far as the background gluon field
$G^{(b)}_{\mu}$ is concerned, we only need to remember that $g
G^{(b)}_{\mu}$ does not get renormalized.

After QCD renormalization, the structure of the effective lagrangian
is the same as in eq.~(\ref{xLeff}), but the Wilson coefficients
$C^Q_i$ are replaced by some other constants that we denote here by
$A^Q_i$. Below, we shall need
\bea
A^Q_j &=& Z_{\psi}^2 \sum_i C^Q_i Z_{ij} \hspace{2cm} \mbox{for  } j=1,2,4,11,
\nonumber\\
A^Q_7 &=& Z_{\psi} Z_g^{-2} \left[       Z_m    \sum_i C^Q_i Z_{i7} 
                                  \;+\; (Z_m-1) \sum_i C^Q_i Z_{i(35)} \right], 
\nonumber\\
A^Q_8 &=& Z_{\psi} Z_g^{-2} \left[       Z_m    \sum_i C^Q_i Z_{i8} 
                                  \;+\; (Z_m-1) \sum_i C^Q_i Z_{i(34)} \right], 
\nonumber\\
A^Q_9 &=& Z_{\psi} Z_g^{-2} \sum_i C^Q_i Z_{i9}.
\label{acoeffs}
\eea

For simplicity, we shall use the $MS$ scheme in the present section.
The $\overline{MS}$ results for the Wilson coefficients will be
obtained later from the $MS$ ones by simply setting $\gamma_E -
\ln(4\pi)$ to zero, i.e. replacing $\kappa$ by $\ln(M_W^2/\mu_0^2)$.

        In the MS scheme, the renormalization constants read 
\bea
Z_g &=& 1 + \f{g^2}{(4 \pi)^2 \e} \left(-\f{1}{2}\beta_0 \right) + {\cal O}(g^4)
\hspace{2cm} \mbox{with  } \beta_0 = \f{23}{3} \mbox{  for 5 active flavours}, 
\nonumber\\ 
Z_m &=& 1 + \f{g^2}{(4 \pi)^2 \e} \left(-\f{1}{2} \gamma^{(0)}_m \right) + {\cal O}(g^4)
\hspace{18mm} \mbox{with  } \gamma_m^{(0)} = 8,
\nonumber\\ 
Z_{\psi} &=& 1 + \f{g^2}{(4 \pi)^2 \e} \left(-\gamma^{(0)}_{\psi} \right) + {\cal O}(g^4)
\hspace{22mm} \mbox{with  } \gamma_{\psi}^{(0)}  = \f{4}{3},
\nonumber\\ 
Z_{ij} &=& \delta_{ij} 
+ \f{g^2}{(4 \pi)^2} \left[ a^{01}_{ij} + \f{1}{\e}   a^{11}_{ij} \right] 
+ \f{g^4}{(4 \pi)^4} \left[ a^{02}_{ij} + \f{1}{\e}   a^{12}_{ij} 
                                        + \f{1}{\e^2} a^{22}_{ij} \right] 
+ {\cal O}(g^6).
\eea
The finite terms $a^{0k}_{ij}$ can be different from zero if and only
if $P_i$ is an evanescent operator and $P_j$ is not. Values of
$a^{0k}_{ij}$ are fixed by requiring that renormalized matrix elements
of evanescent operators vanish in 4 dimensions \cite{evan}. This
requirement is just an extension of the $MS$-scheme definition to
situations where evanescent operators are present.

Our off-shell operator basis is chosen in such a manner that as
many operators as possible are EOM-vanishing. This means that no
linear combination of the remaining operators is EOM-vanishing. In
such a case, the EOM-vanishing operators do not mix into the remaining
ones, i.e. $Z_{ij}=0$ when $P_i$ is EOM-vanishing and $P_j$ is not. In
consequence, we shall need to know explicitly only the mixing among
the physical and evanescent operators.\footnote{
Getting rid of $Z_{i(34)}$ and $Z_{i(35)}$, which enter
eq.~(\ref{acoeffs}), is somewhat tricky -- see subsection 5.4.}

The powers of coupling constants in front of our operators have been
chosen in such a way that terms of order $g^{2n}$ in the
renormalization constants originate from n-loop diagrams in the
effective theory. As one can see, the sum of powers of gauge coupling
constants in front of a given operator is always equal to ``(number of
fields in this operator)--4''. In the original QCD and QED
lagrangians, the powers of coupling constants are equal to ``(number
of fields)--2''. Here, two powers are traded for $G_F$ that normalizes
the effective lagrangian.
        
The renormalization constants are found by calculating ultraviolet
divergent parts of Feynman diagrams in the effective theory. When
doing this, it is essential to clearly separate ultraviolet and
infrared divergences. In order to do so, one can introduce an
auxiliary mass parameter into all the propagator denominators
(including the gluon ones), as explained in ref.~\cite{CMM98.beta}.
All the renormalization constants in the effective theory up to two
loops are known from the former anomalous dimension computations
\cite{M93,BM95,CMM98.df,CMM97} (although some of them need to be
transformed to the ``new'' operator basis (\ref{physical})). Here, we
shall need the one-loop renormalization constant matrix $\hat{a}^{11}$
for $\{P_1, P_2, P_4, P_7, P_8, P_9, P_{11}\}$ only. It reads
\be \label{a11}
\hat{a}^{11} = \left[
\begin{array}{ccccccc}
\vspace{0.2cm}
 * & * &     *    &          0        & 0 & -\f{16}{27} & * \\
\vspace{0.2cm}
 6 & 0 & \f{2}{3} &          0        & 0 & -\f{4}{9}   & 1 \\
\vspace{0.2cm}
 0 & 0 &     *    &          0        & 0 &  \f{16}{27} & 0 \\
\vspace{0.2cm}
 0 & 0 &     0    & \f{16}{3}-\beta_0 & 0 &  0          & 0 \\
\vspace{0.2cm}
 0 & 0 &     0    &      -\f{16}{9}   & * &  0          & 0 \\
\vspace{0.2cm}
 0 & 0 &     0    &          0        & 0 & -\beta_0    & 0 \\
 0 & 0 &     0    &          0        & 0 &     0       & *
\end{array} \right],
\ee
where stars denote non-vanishing entries that are irrelevant for us.

        In addition, for the two-loop matching of photonic penguins in
the charm sector, we shall need
\be \label{a12}
\begin{array}{ccc}
\vspace{0.2cm}
a^{12}_{27}    =  \f{116}{81},  
\hspace{2cm} & 
a^{22}_{27}    =  0, 
\hspace{2cm} & 
a^{01}_{(11)7} = 0,
\\
a^{12}_{29}    = \f{776}{243}, 
\hspace{2cm} & 
a^{22}_{29}    = \f{148}{81},
\hspace{2cm} & 
a^{01}_{(11)9} = \f{64}{27}.  
\end{array}
\ee

        At this point, we are ready to calculate all the necessary 1PI
Green functions on the effective theory side. This turns out to be
very simple, because all the particles in the effective theory are
massless in our approach.\footnote{
Remember that the $b$-quark mass is formally treated here as a
perturbative interaction with an external scalar field, and we include
only terms that are linear in this interaction.}
Consequently, all the loop diagrams vanish in dimensional
regularization, because of the cancellation between ultraviolet and
infrared divergences. In effect, we need to know only the tree-level
matrix element of the effective lagrangian. The ultraviolet
counterterms present in this matrix element reproduce precisely the
infrared divergences in the effective theory, which have to be equal to
the infrared divergences on the SM side. As we shall see, all the
$1/\e^n$ poles will indeed cancel in the matching condition.

External gluons in the Green functions considered on the Standard
Model side have been the background ones. Therefore, we can maintain
only the background gluon field in ${\cal L}_{eff}$, since only
tree-level diagrams are non-vanishing on the effective theory side.
This is why we could omit EOM-vanishing operators proportional to
quantum gluons in our operator basis, even though the calculation is
performed off-shell.

We now write down the effective theory counterparts of the Green
functions considered in subsection 5.2. Their structure follows
directly from tree-level Feynman rules for the operators given in
eqs.~(\ref{physical}) and (\ref{EOM-vanishing}).

        The $b \to s \gamma$ function reads (cf. eq.~(\ref{green1.bsgamma}))
\be \label{eff.green.bsgamma}
i \f{4 G_F}{\sqrt{2}} \f{e P_R}{g^2} 
\left\{ \left( V_{us}^* V_{ub} + V_{cs}^* V_{cb} \right) \sum_{j=1}^{12} \tilde{h}_j S_j
\;\; + \;\;  V_{ts}^* V_{tb} \sum_{j=1}^{12} \tilde{f}_j S_j \right\}
\ee
with the coefficients at the structures $S_2$, $S_8$ and $S_{10}$
given by
\be \label{gamma.act} 
\begin{array}{rclrclrcl}
\tilde{h}_2    &=& -4 A^c_{35}, \hspace{2cm} &
\tilde{h}_8    &=& 2 A^c_{35} - A^c_{36}, \hspace{2cm} &
\tilde{h}_{10} &=& A^c_7 + A^c_{35}, \\[2mm]
\tilde{f}_2    &=& -4 A^t_{35}, &
\tilde{f}_8    &=& 2 A^t_{35} - A^t_{36}, &
\tilde{f}_{10} &=& A^t_7 + A^t_{35}.
\end{array}
\ee
to all orders in QCD. Similarly, for $b \to s\;gluon$ we get 
\be \label{eff.green.bsgluon}
i \f{4 G_F}{\sqrt{2}} \f{P_R T^a}{g} 
\left\{ \left( V_{us}^* V_{ub} + V_{cs}^* V_{cb} \right) \sum_{j=1}^{12} \tilde{u}_j S_j
\;\; + \;\;  V_{ts}^* V_{tb} \sum_{j=1}^{12} \tilde{v}_j S_j \right\}
\ee
with
\be \label{gluon.act}
\begin{array}{rclrclrcl}
\tilde{u}_2    &=& -4 A^c_{34}, \hspace{2cm} &
\tilde{u}_8    &=& 2 A^c_{34} - A^c_{31}, \hspace{2cm} &
\tilde{u}_{10} &=& A^c_8 + A^c_{34},\\[2mm]
\tilde{v}_2    &=& -4 A^t_{34}, &
\tilde{v}_8    &=& 2 A^t_{34} - A^t_{31}, &
\tilde{v}_{10} &=& A^t_8 + A^t_{34}.
\end{array}
\ee

In both the $b \to s \gamma$ and $b \to s\;gluon$ cases, the
coefficients at other structures depend on $A^Q_{32}$ and $A^Q_{33}$,
too. In each of these two cases, coefficients at 12 independent Dirac
structures $S_j$ are given by linear combinations of only 6
independent quantities. It is just a consequence of QCD$\times$QED
gauge invariance of our effective lagrangian. Therefore, the
coefficients at the structures $S_k$ must satisfy $12-6 = 6$ linear
constraints. This must be the case also for the SM Green functions,
because they must match the effective theory ones.  Checking these
constraints on the SM side has been an important cross-check in our
calculation.

        The last function we have to consider on the effective theory
side is the $b \to s c \bar{c}$ one. It takes the form
\bea
&& i \f{4 G_F}{\sqrt{2}} V_{cs}^* V_{cb} 
\left\{       A^c_1 (\gamma_{\mu} P_L T^a) \otimes (\gamma^{\mu} P_L T^a) 
        \;+\; A^c_2 (\gamma_{\mu} P_L    ) \otimes (\gamma^{\mu} P_L    ) 
\right. \nonumber \\ && \left.
\;\; + \;\; A^c_{11}
\left[  (\gamma_{\mu} \gamma_{\nu} \gamma_{\rho} P_L T^a) 
\otimes (\gamma^{\mu} \gamma^{\nu} \gamma^{\rho} P_L T^a)
-16 (\gamma_{\mu} P_L T^a) \otimes (\gamma^{\mu} P_L T^a) \right] \right\} 
\nonumber \\ && 
\; + \; [\mbox{terms proportional to  } (A^Q_{31} + A^Q_4)].
\label{eff.green.bscc}\\[-3mm] \nonumber 
\eea

\noindent
{\large \bf 5.4. The matching}

The Wilson coefficients can be perturbatively expanded as in
eq.~(\ref{expanded.coeffs}). We shall first recover the Wilson
coefficients at all the EOM-non-vanishing operators up to one loop.
Then, two-loop contributions to the coefficients at $P_7$ and $P_9$
will be found.

A careful reader might be surprised that we start the matching without
having considered diagrams with UV counterterms on the SM side. Apart
from the electroweak counterterm proportional to $\bar{s} \slash D b$,
we should include QCD renormalization of the quark wave functions and
masses.

The electroweak counterterm proportional to $\bar{s}\slash D b$ is
taken in the MOM scheme, at $q^2 = 0$ for the $\bar{s} \slash \partial
b$ term, and at vanishing external momenta for the terms containing
gauge bosons. It is achieved by an appropriate flavour-off-diagonal
renormalization of the quark wave functions. The only effect of such a
renormalization in the present case is that the coefficients at the
structure $S_{13}$ in eqs.~(\ref{green1.bsgamma}),
(\ref{green2.bsgamma}) and (\ref{green1.bsgluon}) are completely
renormalized away. This is welcome, because the structure $S_{13}$ was
absent from the effective theory counterparts of these equations
(eqs.~(\ref{eff.green.bsgamma}) and (\ref{eff.green.bsgluon})).

As far as the QCD renormalization of the quark wave functions in
internal lines and in vertices is concerned, it combines to an overall
factor, which could be obtained by renormalizing only those terms in
the vertices that correspond to external fields in a given Green
function.  However, one-loop external quark field renormalization is
the same on the full and effective theory sides.  Consequently, we can
omit counterterms with $Z_{\psi}$ on the SM side and simultaneously
set $Z_{\psi}$ to unity on the effective theory side.

        The same refers to the renormalization of the $b$-quark mass,
since $m_b$ is actually treated as an external scalar field. We omit
the corresponding counterterms on the full theory side and
simultaneously set $Z_m$ to unity on the effective theory side. This
is how we get rid of terms proportional to $(Z_m - 1)$ in
eq.~(\ref{acoeffs}).

        As far as the renormalization of the QCD gauge coupling is
concerned, no such counter\-terms occur on the full theory side in our
particular calculation. On the effective theory side, we maintain all
the necessary factors of $Z_g$.

The last relevant quantity that acquires QCD renormalization on the
full theory side is the top quark mass. However, contributions from
the corresponding counterterm diagrams can be obtained by
differentiating lower order results with respect to $m_t$ (see below).

Let us first match the $b \to s c \bar{c}$ Green function up to one
loop. The first thing to notice is that terms proportional to
$A^Q_{31} + A^Q_4$ in the last line of eq.~(\ref{eff.green.bscc}) are
not important at the considered order, because
\be \label{a4a31}
A^Q_4 = - A^Q_{31} + {\cal O}(g^4).
\ee
The reason for this relation is that the $b \to s d \bar{d}$~ 1PI
Green function acquires its leading contribution only at two loops
in the SM. Lower-order tree-level contributions to this
function must vanish in the effective theory, which implies the above
relation.

Similarly, from the fact that the $b \to s e^+ e^-$~ 1PI function
vanishes at one loop, we find
\be \label{a9a36}
A^Q_9 = + A^Q_{36} + {\cal O}(g^4),
\ee
so long as the $W$-boson boxes and $Z$-boson penguins are not taken
into account on the SM side (as we have assumed at the
very beginning of this section).

        Returning to the $b \to s c \bar{c}$ function, we compare
eqs.~(\ref{green0.bscc}), (\ref{green1.bscc}) and
(\ref{eff.green.bscc}), and immediately find
\bea
A^{c}_1 &=& \f{g^2}{(4 \pi)^2} N^{(1)}_{\e} 
\left( -\f{6}{\e} -15 -\f{39}{2} \e \right) \;+\; {\cal O}(g^4,\e^2),
\nonumber\\[2mm]
A^{c}_2 &=& -1 \;+\; {\cal O}(g^4),
\nonumber\\[2mm]
A^{c}_{11} &=& \f{g^2}{(4 \pi)^2} ( 1 - \e \kappa )
\left( -\f{1}{\e} -\f{3}{2} \right) \;+\; {\cal O}(g^4,\e),
\eea
which implies that (cf. eqs.~(\ref{acoeffs})--(\ref{a11}) with
$Z_{\psi}$ set to unity)
\be
C^{c(0)}_1 = 0, 
\hspace{2cm}
C^{c(0)}_2 = -1,
\hspace{2cm}
C^{c(0)}_{11} = 0,
\ee
and
\bea
C^{c(1)}_1 &=&  N^{(1)}_{\e} 
\left( -\f{6}{\e} -15 -\f{39}{2} \e \right) - \f{1}{\e} C^{c(0)}_2 a^{11}_{21} 
\; + \; {\cal O}(\e^2) \nonumber\\[2mm]
&=& -15 + 6 \kappa \;+\; \e \left(-\f{39}{2}+15\kappa-3\kappa^2-\f{1}{2}\pi^2\right)
+ {\cal O}(\e^2),
\\
C^{c(1)}_2 &=& 0,
\\
C^{c(1)}_{11} &=&  
( 1 - \e \kappa ) \left( -\f{1}{\e} -\f{3}{2} \right) 
- \f{1}{\e} C^{c(0)}_2 a^{11}_{2(11)} \;+\; {\cal O}(\e) \nonumber\\
&=& -\f{3}{2} + \kappa \;+\; {\cal O}(\e).
\eea
Indeed, all the $1/\e$ poles have cancelled in the final results for
the one-loop Wilson coefficients.

        The coefficient $C^c_2$ is the only one that acquires a
tree-level contribution in our calculation. For all the other
coefficients considered below, we have $C^{Q(0)}_i=0$.

Let us now turn to the $b \to s\;gluon$ matching. Comparing
eqs.~(\ref{green1.bsgluon})\footnote{
  Without $S_{13}$, since it has been renormalized away by the
  electroweak counterterm mentioned in the beginning of this
  subsection.}
and (\ref{eff.green.bsgluon}), and solving the trivial set of linear
equations \{(\ref{gluon.act}),(\ref{a4a31})\}, one finds
\bea
A^c_4 &=& \f{g^2}{(4 \pi)^2} N^{(1)}_{\e} 
\left( \f{1}{2} u^{(1)}_2 + u^{(1)}_8 \right) + {\cal O}(g^4,\e^2),
\nonumber\\
A^c_8 &=& \f{g^2}{(4 \pi)^2} (1-\e\kappa)
\left( \f{1}{4} u^{(1)}_2 + u^{(1)}_{10} \right) + {\cal O}(g^4,\e^2),
\eea
which implies that (cf. eqs.~(\ref{acoeffs})--(\ref{a11}))
\bea
C^{c(1)}_4 &=& N^{(1)}_{\e} 
\left( \f{1}{2} u^{(1)}_2 + u^{(1)}_8 \right) - \f{1}{\e} a^{11}_{24} C^{c(0)}_2
+ {\cal O}(\e^2)
\nonumber\\[2mm]
&=& \f{7}{9} + \f{2}{3} \kappa + 
\e \left(\f{77}{54}-\f{7}{9}\kappa-\f{1}{3}\kappa^2-\f{1}{18}\pi^2 \right)
+ {\cal O}(\e^2),
\nonumber\\[2mm]
C^{c(1)}_8 &=& (1-\e \kappa) \left( \f{1}{4} u^{(1)}_2 + u^{(1)}_{10} \right) 
+ {\cal O}(\e^2)
\nonumber\\[2mm]
&=& \f{1}{3} + \e \left(\f{11}{18}-\f{1}{3}\kappa \right) + {\cal O}(\e^2).
\eea
Similarly, \vspace{-4mm}
\bea
C^{t(1)}_4 &=& 
(1-\e\kappa) \left( \f{1}{2} v^{(1)}_2(x) + v^{(1)}_8(x)    \right) 
+ {\cal O}(\e^2),
\nonumber\\[2mm]
C^{t(1)}_8 &=& 
(1-\e\kappa) \left( \f{1}{4} v^{(1)}_2(x) + v^{(1)}_{10}(x) \right) 
+ {\cal O}(\e^2).
\eea

        Finally, we perform the $b \to s \gamma$ matching. Comparing
eqs.~(\ref{green1.bsgamma}), (\ref{green2.bsgamma}) and
(\ref{eff.green.bsgamma}), and solving the trivial set of linear
equations \{(\ref{gamma.act}),(\ref{a9a36})\}, one finds
\bea
A^c_7 &=& \f{g^2}{(4 \pi)^2} \left[ (1-\e\kappa)
\left( \f{1}{4} h^{(1)}_2 + h^{(1)}_{10} \right) + {\cal O}(\e^2) \right]
\nonumber\\
&+& \f{g^4}{(4 \pi)^4} \left[ (1-2\e\kappa)
\left( \f{1}{4} h^{(2)}_2 + h^{(2)}_{10} \right) + {\cal O}(\e) \right]
+ {\cal O}(g^6),
\nonumber\\
A^c_9 &=& \f{g^2}{(4 \pi)^2} \left[ N^{(1)}_{\e} 
\left( -\f{1}{2} h^{(1)}_2 - h^{(1)}_8 \right) + {\cal O}(\e^2) \right]
\nonumber\\
&+& \f{g^4}{(4 \pi)^4} \left[ N^{(2)}_{\e} 
\left( -\f{1}{2} h^{(2)}_2 - h^{(2)}_8 \right) + {\cal O}(\e) \right]
+ {\cal O}(g^6),
\eea
which implies that (cf. eqs.~(\ref{acoeffs})--(\ref{a12}) with
$Z_{\psi}$ and $Z_m$ set to unity)
\bea
C^{c(1)}_7 &=& (1-\e \kappa) \left( \f{1}{4} h^{(1)}_2 + h^{(1)}_{10} \right) 
+ {\cal O}(\e^2)
\nonumber\\[2mm]
&=& \f{23}{36} + \e \left(\f{145}{216}-\f{23}{36}\kappa \right) + {\cal O}(\e^2),
\nonumber\\[2mm]
C^{c(1)}_9 &=& N^{(1)}_{\e} 
\left( -\f{1}{2} h^{(1)}_2 - h^{(1)}_8 \right) - \f{1}{\e} a^{11}_{29} C^{c(0)}_2
+ {\cal O}(\e^2)
\nonumber\\[2mm]
&=& -\f{38}{27} -\f{4}{9} \kappa + 
\e \left(-\f{247}{162}+\f{38}{27}\kappa+\f{2}{9}\kappa^2+\f{1}{27}\pi^2 \right)
+ {\cal O}(\e^2),
\eea
and
\bea
C^{c(2)}_7 &=& (1-2\e \kappa) \left( \f{1}{4} h^{(2)}_2 + h^{(2)}_{10} \right) 
-\f{1}{\e} \left[ a^{12}_{27} C^{c(0)}_2 + (a^{11}_{77} + \beta_0) C^{c(1)}_7
                + a^{11}_{87} C^{c(1)}_8 \right] + {\cal O}(\e)
\nonumber\\[2mm]
&=& (1-2\e \kappa) \left( \f{112}{81\e} -\f{107}{243} \right) - \f{116}{81\e} (-1) 
\nonumber\\[2mm] && 
- \f{16}{3\e} \left( \f{23}{36} + \f{145\e}{216}-\f{23\e}{36}\kappa \right) 
+ \f{16}{9\e} \left( \f{1}{3} + \f{11\e}{18}-\f{\e}{3}\kappa \right) + {\cal O}(\e)
\nonumber\\[2mm]
&=& -\f{713}{243} + \f{4}{81} \kappa + {\cal O}(\e),
\\[4mm]
C^{c(2)}_9 &=& N^{(2)}_{\e} 
\left( -\f{1}{2} h^{(2)}_2 - h^{(2)}_8 \right) 
-\f{1}{\e^2} \left( a^{22}_{29} + \beta_0 a^{11}_{29} \right) C^{c(0)}_2 
\nonumber\\[2mm] &&
-\f{1}{\e} \left[ a^{12}_{29} C^{c(0)}_2 + a^{11}_{19} C^{c(1)}_1
  + a^{11}_{49} C^{c(1)}_4 \right] - a^{01}_{(11)9} C^{c(1)}_{11} + {\cal O}(\e)
\nonumber\\[2mm] 
&=& \left[ 1 - 2 \e \kappa + \e^2 \left( \f{\pi^2}{6} + 2\kappa^2 \right) \right] 
\left( \f{128}{81\e^2} +\f{1496}{243\e} + \f{5924}{729} +\f{128}{243}\pi^2 \right) 
\nonumber\\[2mm] &&
+ \f{128}{81\e^2} (-1)  - \f{776}{243\e} (-1) 
+ \f{16}{27\e} \left[-15+6\kappa+\e\left(-\f{39}{2}+15\kappa-3\kappa^2-\f{1}{2}\pi^2\right)\right]
\nonumber\\[2mm] &&
- \f{16}{27\e} \left[\f{7}{9}+\f{2}{3}\kappa
    +\e\left(\f{77}{54}-\f{7}{9}\kappa-\f{1}{3}\kappa^2-\f{1}{18}\pi^2\right)\right]
-\f{64}{27} \left( -\f{3}{2} + \kappa \right) + {\cal O}(\e)
\nonumber\\[2mm] 
&=& -\f{524}{729} -\f{16}{3}\kappa +\f{128}{81}\kappa^2 +\f{128}{243}\pi^2 +{\cal O}(\e).
\eea
Again, all the $1/\e^n$ poles have cancelled in the final results.
        
        Similarly, in the top sector we find
\bea
C^{t(1)}_7 &=& (1-\e\kappa) 
\left[ \f{1}{4} f^{(1)}_2(x) + f^{(1)}_{10}(x) \right] + {\cal O}(\e^2),
\nonumber\\ 
C^{t(1)}_9 &=& (1-\e\kappa) 
\left[ -\f{1}{2} f^{(1)}_2(x) - f^{(1)}_8(x) \right] + {\cal O}(\e^2),
\nonumber\\ 
C^{t(2)}_7 &=& (1-2\e\kappa) 
\left[ \f{1}{4} f^{(2)}_2(x) + f^{(2)}_{10}(x) \right] 
-\f{1}{\e} \gamma_m^{(0)} x \f{\partial}{\partial x} C^{t(1)}_7
-\f{1}{\e} \left[ (a^{11}_{77} + \beta_0) C^{t(1)}_7 + a^{11}_{87} C^{t(1)}_8 \right]
+ {\cal O}(\e),
\nonumber\\ 
C^{t(2)}_9 &=& (1-2\e\kappa) 
\left[ -\f{1}{2} f^{(2)}_2(x) - f^{(2)}_8(x) \right] 
-\f{1}{\e} \gamma_m^{(0)} x \f{\partial}{\partial x} C^{t(1)}_9
-\f{1}{\e} a^{11}_{49} C^{t(1)}_4 + {\cal O}(\e).
\eea
Here, the $x$-derivative terms stand for contributions from the
top-quark mass renormalization on the full theory side. Instead of
including these terms, we could just calculate the corresponding
one-loop SM diagrams with counterterm insertions. However, derivatives
give us the same results much faster.

        It is easy to verify that all the $1/\e$ poles indeed
cancel in $C^{t(2)}_7$ and $C^{t(2)}_9$. As usual, the ${\cal O}(\e)$
parts of the one-loop Wilson coefficients have affected the results of
the two-loop matching.

The results for $C^{c(0)}_2$, $C^{c(1)}_1$, $C^{c(1)}_2$,
$C^{Q(1)}_4$, $C^{Q(1)}_7$, $C^{Q(1)}_9$, $C^{Q(2)}_7$ and
$C^{Q(2)}_9$ obtained in the present section have already been 
summarized in section~\ref{matching}, after passing to the
$\overline{MS}$ scheme, i.e. replacing $\kappa$ by $\ln(M_W^2/\mu_0^2)$.
All the other matching conditions summarized there have been found
in an analogous manner.  In the two-loop $Z$-penguin contributions to
$C^Q_9$ and $C^Q_{10}$, the effect of renormalizing the $\bar{s}
\slash D b$ term on the SM side was less trivial than in this section.
In the two-loop matching for $P^c_1$ and $P^c_2$, some care was
required at renormalizing the top-quark loop contributions in the MOM
scheme.  In addition, scalar integrals with three non-vanishing masses
were necessary \cite{DT93}.  Nevertheless, the basic algorithm
remained the same as in the $P_7$ and $P_9$ cases, which we have
described in detail here.

\ \\
{\bf \Large Summary}

We have evaluated two-loop matching conditions for all the operators
relevant to \linebreak $B \to X_s l^+ l^-$ in the SM. Details of this
calculation have been presented only for the operator $P_7$ and for
the photonic penguin contribution to the operator $P_9$. As far as the
remaining matching conditions are concerned, only the final results
have been given. However, the method of the calculation was very
similar in all the considered cases.

Our results allowed to remove an important ($\sim \pm 16\%$)
uncertainty due to the matching scale $\mu_0$ from the prediction for
$BR[B \to X_s l^+ l^-]$ for low invariant mass of the emitted lepton
pair ($\s \in [0.05,0.25]$). The obtained Standard Model prediction
for the branching ratio integrated over this domain is $1.46 \times
10^{-6}$.  This result would change to $2.92\times10^{-6}$ if the
Wilson coefficient $\tilde{C}_7^{eff}(\mu_b)$ had an opposite sign, as
it might happen in certain extensions of the SM.

There remains a sizeable ($\sim \pm 13\%$) perturbative uncertainty in
the above SM result, which is due to the unknown two-loop matrix
elements of the four-quark operators. Calculable non-perturbative
effects which have been included in our result are smaller than this
uncertainty.  Estimates of other non-perturbative effects suggest that
they are not larger. Therefore, the next step in improving the
accuracy of the theoretical prediction should be a calculation of the
two-loop matrix elements of the four-fermion operators and one-loop
matrix elements of the ``magnetic moment'' ones.

\ \\
{\bf \Large Acknowledgements}

We are grateful to Andrzej Buras for suggesting us to work at the
matching-scale dependence of $B \to X_s l^+ l^-$.  C.B and J.U. thank
Frank Krauss, Klaus Schubert and Gerhard Soff, while M.M. thanks
Patricia Ball, Gerhard Buchalla, Paolo Gambino and Frank Kr\"uger for
helpful discussions.

This work has been supported in part by the German Bundesministerium
f{\"u}r Bildung und Forschung under contracts 06~DD~823, 05 HT9WOA
(J.U.) and 06~TM~874 (M.M). M.M. has been supported in part by the DFG
project Li~519/2-2, as well as by the Polish Committee for Scientific
Research under grant 2~P03B~014~14, 1998-2000.

\ \\
{\bf \Large Appendix}

Here, we give the eight evanescent operators that were used in
evaluating the anomalous dimension matrices given in
section~\ref{coefficients}. Their explicit form defines what the $MS$
scheme means in the effective theory. As before, the symbol $Q$ stands
either for $u$ or for $c$.
\bea 
P^Q_{11} &=& (\bar{s}_L \gamma_{\mu_1}
                        \gamma_{\mu_2}
                        \gamma_{\mu_3} T^a Q_L)(\bar{Q}_L \gamma^{\mu_1} 
                                                          \gamma^{\mu_2}
                                                          \gamma^{\mu_3} T^a b_L) 
-16 P^Q_1, \nonumber \\
P^Q_{12} &=& (\bar{s}_L \gamma_{\mu_1}
                        \gamma_{\mu_2}
                        \gamma_{\mu_3}     Q_L)(\bar{Q}_L \gamma^{\mu_1} 
                                                          \gamma^{\mu_2}
                                                          \gamma^{\mu_3}     b_L) 
-16 P^Q_2, \nonumber \\
P_{15} &=& (\bar{s}_L \gamma_{\mu_1}
                      \gamma_{\mu_2}
                      \gamma_{\mu_3}
                      \gamma_{\mu_4}
                      \gamma_{\mu_5}     b_L)\sum_q(\bar{q} \gamma^{\mu_1} 
                                                            \gamma^{\mu_2}
                                                            \gamma^{\mu_3}
                                                            \gamma^{\mu_4}
                                                            \gamma^{\mu_5}     q) 
-20 P_5 + 64 P_3, \nonumber \\
P_{16} &=& (\bar{s}_L \gamma_{\mu_1}
                      \gamma_{\mu_2}
                      \gamma_{\mu_3}
                      \gamma_{\mu_4}
                      \gamma_{\mu_5} T^a b_L)\sum_q(\bar{q} \gamma^{\mu_1} 
                                                            \gamma^{\mu_2}
                                                            \gamma^{\mu_3}
                                                            \gamma^{\mu_4}
                                                            \gamma^{\mu_5} T^a q) 
-20 P_6 + 64 P_4, \nonumber \\
P^Q_{21} &=& (\bar{s}_L \gamma_{\mu_1}
                        \gamma_{\mu_2}
                        \gamma_{\mu_3}
                        \gamma_{\mu_4}
                        \gamma_{\mu_5} T^a Q_L)(\bar{Q}_L \gamma^{\mu_1} 
                                                          \gamma^{\mu_2}
                                                          \gamma^{\mu_3}
                                                          \gamma^{\mu_4}
                                                          \gamma^{\mu_5} T^a b_L) 
-20 P^Q_{11} - 256 P^Q_1, \\
P^Q_{22} &=& (\bar{s}_L \gamma_{\mu_1}
                        \gamma_{\mu_2}
                        \gamma_{\mu_3}
                        \gamma_{\mu_4}
                        \gamma_{\mu_5} T   Q_L)(\bar{Q}_L \gamma^{\mu_1} 
                                                          \gamma^{\mu_2}
                                                          \gamma^{\mu_3}
                                                          \gamma^{\mu_4}
                                                          \gamma^{\mu_5}     b_L) 
-20 P^Q_{12} - 256 P^Q_2, \nonumber \\
P_{25} &=& (\bar{s}_L \gamma_{\mu_1}
                      \gamma_{\mu_2}
                      \gamma_{\mu_3}
                      \gamma_{\mu_4}
                      \gamma_{\mu_5}
                      \gamma_{\mu_6}
                      \gamma_{\mu_7}     b_L)\sum_q(\bar{q} \gamma^{\mu_1} 
                                                            \gamma^{\mu_2}
                                                            \gamma^{\mu_3}
                                                            \gamma^{\mu_4}
                                                            \gamma^{\mu_5}
                                                            \gamma^{\mu_6}
                                                            \gamma^{\mu_7}     q) 
-336 P_5 + 1280 P_3, \nonumber \\
P_{26} &=& (\bar{s}_L \gamma_{\mu_1}
                      \gamma_{\mu_2}
                      \gamma_{\mu_3}
                      \gamma_{\mu_4}
                      \gamma_{\mu_5}
                      \gamma_{\mu_6}
                      \gamma_{\mu_7} T^a b_L)\sum_q(\bar{q} \gamma^{\mu_1} 
                                                            \gamma^{\mu_2}
                                                            \gamma^{\mu_3}
                                                            \gamma^{\mu_4}
                                                            \gamma^{\mu_5}
                                                            \gamma^{\mu_6}
                                                            \gamma^{\mu_7} T^a q) 
-336 P_6 + 1280 P_4. \nonumber 
\eea

\setlength {\baselineskip}{0.2in}
 

\begin{thebibliography}{99}
\newcommand{\np}[3]{Nucl. Phys. {\bf B#1} (#2) #3}
\newcommand{\pl}[3]{Phys. Lett. {\bf B#1} (#2) #3}
\newcommand{\pr}[3]{Phys. Rev.  {\bf D#1} (#2) #3}
\newcommand{\prl}[3]{Phys. Rev. Lett. {\bf #1} (#2) #3}
\newcommand{\prp}[3]{Phys. Rept. {\bf #1} (#2) #3}
\newcommand{\zpc}[3]{Z. Phys. {\bf C#1} (#2) #3}

\bibitem{LW96} Z.~Ligeti and M.B.~Wise, \pr{53}{1996}{4937}.
\bibitem{FLS94} A.F.~Falk, M.~Luke and M.J.~Savage, \pr{49}{1994}{3367}.
\bibitem{AHHM97} A.~Ali, G.~Hiller, L.T.~Handoko and T.~Morozumi, \pr{55}{1997}{4105}.
\bibitem{CRS97} J-W.~Chen, G.~Rupak and M.J.~Savage, \pl{410}{1997}{285}.
\bibitem{BIR98} G.~Buchalla, G.~Isidori and S.-J.~Rey, \np{511}{1998}{594}.
\bibitem{BI98} G.~Buchalla and G.~Isidori, \np{525}{1998}{333}.
\bibitem{M93} M.~Misiak, \np{393}{1993}{23}, {\bf B439} (1995) 461 (E).
\bibitem{BM95} A.J.~Buras and M.~M{\"u}nz, \pr{52}{1995}{186}.
\bibitem{CMM98.df} K.~Chetyrkin, M.~Misiak and M.~M{\"u}nz, \np{520}{1998}{279}.
\bibitem{2mtch}
  K.~Adel and Y.P.~Yao, \pr{49}{1994}{4945};\\
  C.~Greub and T.~Hurth, \pr{56}{1997}{2934};\\
  A.J.~Buras, A.~Kwiatkowski and N.~Pott, \np{517}{1998}{353};\\
  M.~Ciuchini, G.~Degrassi, P.~Gambino and G.F.~Giudice, \np{527}{98}{21};\\
  C.~Bobeth, M.~Misiak and J.~Urban, hep-ph/9904413, to appear in Nucl. Phys. {\bf B}.
\bibitem{MU99} M.~Misiak and J.~Urban, \pl{451}{1999}{161}.
\bibitem{BB93} G.~Buchalla and A.J.~Buras, \np{400}{1993}{225}.
\bibitem{BB99} G.~Buchalla and A.J.~Buras, \np{548}{1999}{309}.
\bibitem{CMM97} K.~Chetyrkin, M.~Misiak and M.~M{\"u}nz, \pl{400}{1997}{206},\\ 
                                                       {\bf B425} (1998) 414 (E).
\bibitem{PData98} Particle Data Group, Eur. Phys. J. {\bf C3} (1998) 1.
\bibitem{KS96} F.~Kr\"uger and L.M.~Sehgal, \pl{380}{1996}199.
\bibitem{CM78} N. Cabibbo and L. Maiani, \pl{79}{1978}{109}.
\bibitem{N89} Y. Nir, \pl{221}{1989}{184}.
\bibitem{BaBu99} C.W.~Bauer and C.N.~Burrell, hep-ph/9907517.
\bibitem{AH98} A.~Ali and G.~Hiller, \pr{58}{1998}{074001};\\
               G.~Hiller, hep-ph/9809505.
\bibitem{CMW96} P.~Cho, M.~Misiak and D.~Wyler, \pr{54}{1996}{3329}.
\bibitem{KBD90} 
  J.~K{\"u}blbeck, M.~B{\"o}hm and A.~Denner, Comput. Phys. Commun. 60 (1990) 165.
\bibitem{DT93} A.I.~Davydychev and J.B.~Tausk, \np{397}{1993}{123}.
\bibitem{evan} 
  A.J.~Buras and P.H.~Weisz, \np{333}{1990}{66};\\
  M.~J.~Dugan and B.~Grinstein, \pl{256}{1991}{239};\\
  S.~Herrlich and U.~Nierste, \np{455}{1995}{39}.
\bibitem{CMM98.beta} K.~Chetyrkin, M.~Misiak and M.~M{\"u}nz, \np{518}{1998}{473}.
\end{thebibliography}
\end{document}